\documentclass[12pt]{iopart}
\newcommand{\PeM}{{\rm PM}}
\newcommand{\MM}{{\rm MM}}
\newcommand{\bulk}{{\rm bulk}}
\newcommand{\Sfold}{{\rm Sfold}}
\newcommand{\Pfold}{{\rm Pfold}}
\newcommand{\NS}{{\rm NS}}
\newcommand{\spec}{{\rm S}}
\usepackage{graphicx}

\begin{document}

\title[The physics of oligonucleotide microarrays]{Understanding the physics of oligonucleotide microarrays: the Affymetrix spike-in data reanalysed}  

\author{Conrad J.\ Burden$^1$}
\address{$^1$Centre for Bioinformation Science\\
      John Curtin School of Medical Research and Mathematical Sciences Institute\\
      Australian National University,
      Canberra, ACT 0200, Australia}
\ead{Conrad.Burden@anu.edu.au}

\begin{abstract}
The Affymetrix U95 and U133 Latin Square spike-in datasets are reanalysed, together with a dataset from a version of the U95 spike-in experiment without a complex non-specific background.  The approach uses a physico-chemical model which includes the effects the specific and non-specific hybridisation and probe folding at the microarray surface, target folding and hybridisation in the bulk RNA target solution, and duplex dissociation during the post-hybridisatoin washing phase.  The model predicts a three parameter hyperbolic response function that fits well with fluorescence intensity data from all three datasets.  The importance of the various hybridisation and washing effects in determining each of the three parameters is examined, and some guidance is given as to how a practical algorithm for determining specific target concentrations might be developed.  
\end{abstract}

\pacs{87.15.-v, 82.39.Pj}

\maketitle

\section{Introduction}
\label{sec:Intro}

A number of papers~\cite{Hekstra03, Held03, Burden04, Binder04, Binder06b, Carlon06, Heim06, Burden06,Held06} have used chemical adsorption models to analyse data from two well-known Affymetrix Latin-Square spike-in experiments~\cite{AffyLatinSq}, known as the U95 and U133 datasets.   The immediate aim of these papers has been to study the physical processes responsible for converting concentrations of specific target RNA in prepared solutions hybridised onto microarrays to measured fluorescence intensities.  Their ultimate aim has been to provide biologists with a practical algorithm for estimating absolute specific target concentrations in the presence of a complex non-specific background from fluorescence intensity data.  Even though there are a number of existing algorithms (or ``expression measures'') of varying degees of statistical sophistication currently available to and widely used by biologists, such algorithms pay little attention to the detailed physics and chemistry of hybridisation at the microarray surface.  As a result they are prone to underestimating fold changes at high target concentrations because saturation effects are not modelled, and at low target concentrations because of a failure to account adequately for non-specific hybridisation~\cite{Burden07}.  Furthemore, the measures reported are not target concentrations as such, but surrogates derived directly from fluorescence intensities.  At best, they could be described as an unknown increasing function of specific target concentrations, modulo statistical noise.  

Before a reliable algorithm for inferring target concentrations can be developed, an accurate model of the physics and chemistry of the process is needed.  In a recent review of chemical adsorption effects at the microarray surface~\cite{Binder06a}, Binder has described in detail a number of processes influencing fluorescence intensity measurements, including bulk dimerisation of target molecules in solution, non-specific hybridisation and probe folding at the microarray surface, and partial zippering of duplexes.  Analyses of the Affymetrix spike-in data have provided strong evidence that these effects cannot be ignored.  For instance, Binder and Preibisch~\cite{Binder06b} have isolated the effects of specific and non-specific binding, Carlon and Heim\cite{Carlon06} and Heim  et al.~\cite{Heim06} have stressed the importance of bulk hybridisation and target folding in solution, and by analysing other data sets, Matveeva et al.\cite{Matveeva03} have produced evidence for the importance of probe folding.  

More recently, analyses of the U95 dataset~\cite{Burden06,Held06} and subsequent experimental evidence~\cite{Skvortsov07} have demonstrated the significant influence of the post-hybridisation washing step in determining fluorescence intensities.  Although the washing step has an important scaling effect over the whole range of target concentrations, it is particularly noticeable as a probe sequence dependent scaling of the asymptote of the measured intensity isotherm at saturation concentrations.  Because adsorption theories of the hybridisation step which neglect the post-hybridisation washing predict a common asymptote for these isotherms, many adsorption-model-based analyses of the Affymetrix Latin Square data sets have forced the data to fit a common asymptote~\cite{Held03, Carlon06, Heim06, Binder06b}, in spite of strong statistical evidence to the contrary~\cite{Burden04,Hekstra03}.  

In this paper we reanalyse both the U95 and U133 datasets, and a third dataset which is analogous to the U95 dataset, but without the complex background.  Our analysis uses a chemical adsorption model which includes a number of chemical reactions occurring during the hybridisation and post-hybridisation washing steps.   Details of the datasets are summarised in Section~\ref{sec:Datasets}, and all three datasets are shown to be consistent with the empirical observations previously observed for the U95 dataset, namely that measured fluorescence intensities follow a hyperbolic isotherm with probe-sequence dependent parameters~\cite{Burden04}.  Our physico-chemical model is described in Section~\ref{sec:Model} and demonstrated to be quantitatively consistent with these observations in Section~\ref{sec:Qualitative}. A quantitative analysis of how well the model fits the three datasets is given in Section~\ref{sec:Quantitative}.  An analysis in Section~\ref{sec:Prediction} gives some indication of how intensity measurements over a whole microarray might be used to infer some of the isotherm parameters for each feature on the microarray, which in turn has the potential to contribute towards development of a practical algorithm for estimating target concentrations.  Concluding comments are given in Section~\ref{sec:Conclusions}.  

\section{Datasets and empirical fits}
\label{sec:Datasets}

Three datasets are analysed in this paper, the two publicly available Affymetrix Latin-Square spike-in experiments~\cite{AffyLatinSq}, known as the U95 and U133 datasets, and a version of the U95 dataset without the complex human pancreas background, kindly made available to us by Affymetrix.   We will refer to these datasets as numbers I, II and III respectively.  

In the manufacture of Affymetrix GeneChip\textregistered ~arrays, single strand DNA probes, 25 bases in length are 
synthesized base by base onto a quartz substrate using a photolithographic process.  They are 
attached to the substrate via short covalently bonded linker molecules roughly 10 nanometres apart.  
A microarray chip surface is divided into some hundreds of thousands of regions called features, commonly 11 to 20 microns square, and with the single strand DNA probes within each feature being synthesized to a specific nucleotide sequence.  

The main step in the laboratory process of gene detection with microarrays is the 
hybridization of cRNA target molecules, fractionated to lengths of typically 50 to 200 bases and with biotin labels attached to their U and C bases, onto the single strand DNA probes.  The hybridisation is performed at $45^\circ$C for 16 hours.  The microarray is then washed to remove excess cRNA target, the biotin labels stained with fluorescent dye, and the density of hybridized probe-target duplexes in each feature detected via intensity measurements of the fluorescent dye.  
Each gene or EST is represented by a set of pairs of features (16 pairs in the case of the U95 dataset and 11 pairs for U133) using sequences of length 25 selected for their predicted hybridization properties and 
specificity to the target gene.  The first element of the pair, termed the perfect match (PM), 
is designed to be an exact match to the target sequence, while the second element, the 
mismatch (MM), is identical except for the middle (13th) base being replaced by its complement.  

In the Affymetrix spike-in experiments, RNA transcripts were spiked in at cyclic permutations of a set of known concentrations together with a complex background of cRNA extracted from human pancreas (dataset I), human adenocarcinoma cell line (dataset II), or no background (dataset III).  Each of datasets I and III consist of PM and MM fluorescence intensity values from a set of 14 probe sets corresponding to 14 separate genes, each containing 16 probe pairs.  For each probe set a set of fluorescence intensity values was obtained for the 14 spiked-in concentrations 0, 0.25, 0.5, 1,2, \ldots, 1024 pM.  In common with previous analyses of dataset I, the following analysis of datasets I and III is restricted to 12 of the 14 genes, omitting data from the two defective probesets, 36889\_at and 407\_at.  Dataset II consists of PM and MM fluorescence intensity values from a set of 38 probe sets corresponding to 38 separate genes, each containing 11 probe pairs.  For each probe set a set of fluorescence intensity values was obtained for the 14 spiked-in concentrations 0, 0.125, 0.25, 0.5, 1, \ldots, 512 pM.  Dataset II also contains data from a further 4 bacterial gene probe sets each containing 20 probe pairs, which we do not include in the current analysis.  

In a previous paper~\cite{Burden04} it was demonstrated that the dataset I is consistent with the empirical observation that the measured fluorescence intensities $I(x)$ at spike-in concentration $x$ are sampled from a Gamma distribution with constant coefficient of variation about a mean given by a hyperbolic response curves of the form
\begin{equation}
I(x) = A + B\frac{Kx}{1 + Kx}.     \label{hyperbolicResponse}
\end{equation}
Importantly, it was further shown using a thorough statistical analysis that each of the parameters $A$, $B$ and $K$ is feature (i.e. probe-sequence) dependent, and that the asymptote,  $\lim_{x\rightarrow\infty}I(x) = A + B$, is almost invariably lower for the MM feature than the neighbouring perfect-match PM feature within any PM/MM pair.  

On the assumption that by far the greatest proportion of the intensity signal across a whole microarray in either dataset I or II comes from the complex background, we have preprocessed the data by carrying out a quantile normalisation across each of these two datasets~\cite{Irizarry03}.  Thus all microarrays within a particular dataset have a common distribution of fluorescence intensities after preprocessing.  Because of the absence of a complex background, we have not carried out this step for dataset III. Instead we have included in the fit the ``wafer dependent scaling" described as Model {\bf E} in ref.~\cite{Burden04} to allow for the fact that the three replicates of the experiment used microarrays constructed from three separate wafers.  That is to say, we fit to the data a model of the form $I(x) = \lambda_w[A_p + B_p K_p x/(1 + K_p x)]$, where the index $w = 1, 2, 3$ labels the three replicates, the index $p$ labels a feature within a probe set and the scaling factors satisfy $\frac{1}{3}\sum_w \lambda_w = 1$.  

\begin{figure}
\includegraphics[scale=0.9]{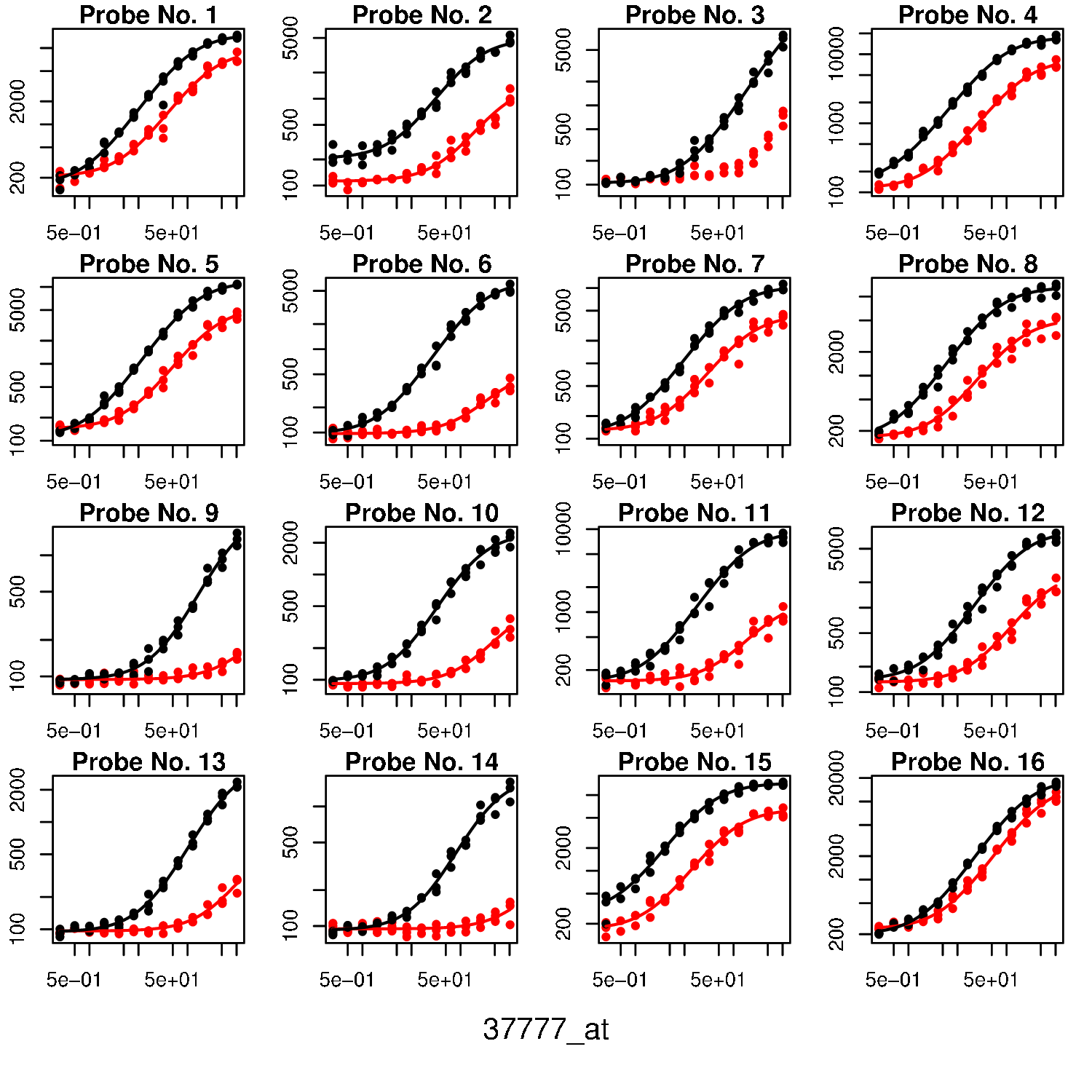}
\centering
\caption{Fits of fluorescence intensity data to the hyperbolic response curve Eq.~\ref{hyperbolicResponse} for the spiked-in transcript 37777\_at for dataset I.  PM data are in blcak and MM data in red.  Spike-in concentrations (horizontal axes) in picomolar on a logarithmic scale, and fluorescence intensities after quantile normalisation (vertical axes) are in the arbitrary units used in Affymetrix cel files on a logarithmic scale.}
\label{fig:fit37777}
\end{figure}

\begin{figure}
\includegraphics[scale=0.9]{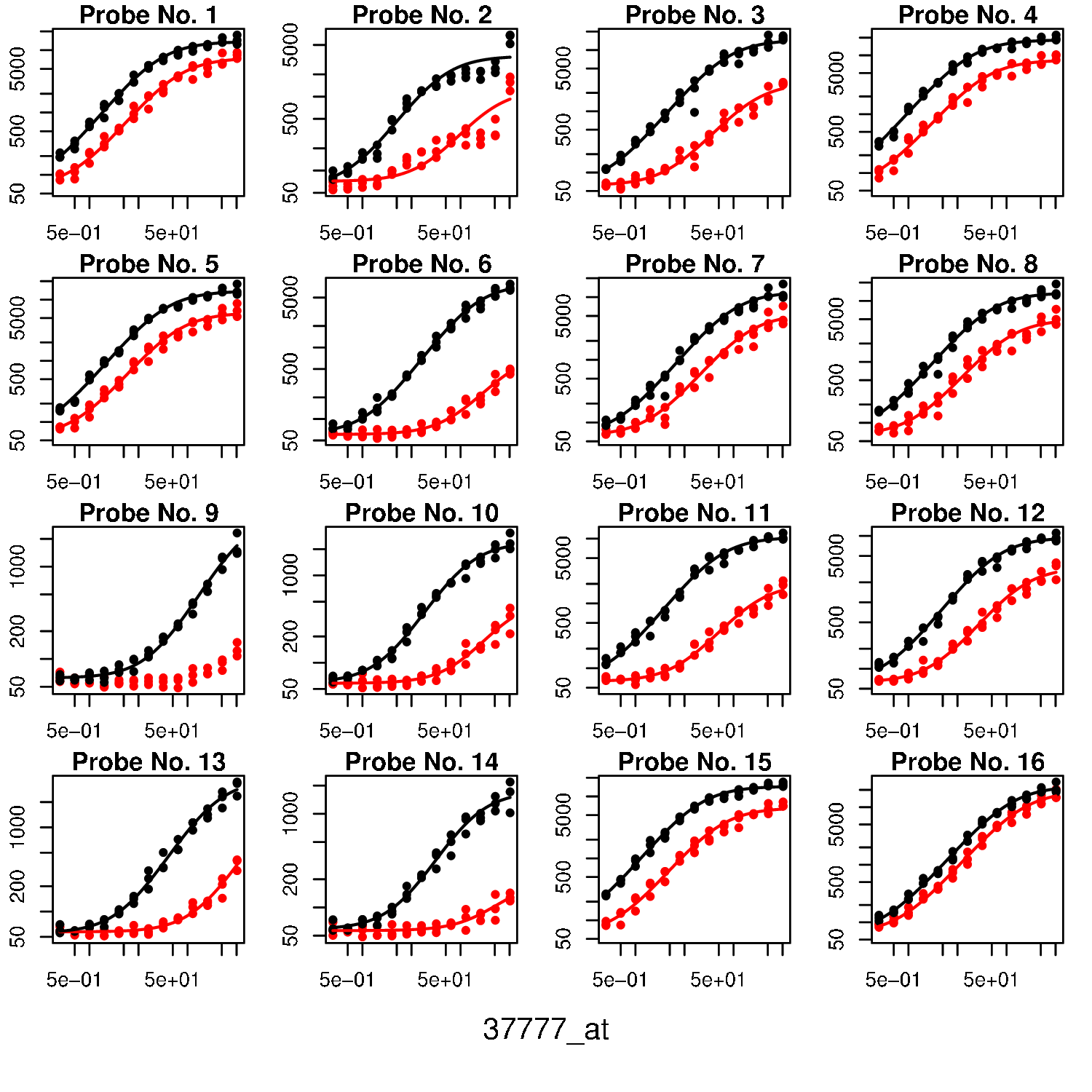}
\centering
\caption{Fits of fluorescence intensity data to the hyperbolic response curve Eq.~\ref{hyperbolicResponse} for the spiked-in transcript 37777\_at for dataset III, without the complex human pancreas background.}
\label{fig:fit37777nbg}
\end{figure}

In Figures~\ref{fig:fit37777} and \ref{fig:fit37777nbg} are plotted fits of fluorescence intensity data to the response curve Eq.~\ref{hyperbolicResponse} for one of the spiked-in transcripts for datasets I and III.  A complete set of analogous plots for all transcripts from all three datasets can be found at {\tt http://dayhoff.anu.edu.au/\~{}conrad/Spike\_in\_Isotherms/}.  

For a small minority of features the fit gives negative values to some of the parameters $A$, $B$ or $K$, whereas the physical model set out in Section~\ref{sec:Model} predicts that all three parameter should be positive.  This problem is more slightly more prevalent for MM features than for PM features, and is most acute for dataset II.  In some cases, such as probe number 3 in Fig.~\ref{fig:fit37777}, it appears that the data does not extend far enough into the high concentration, i.e. saturation, limit to allow an acceptable fit.  In other cases, such as probe number 9 of Fig.~\ref{fig:fit37777nbg}, the data may be too noisy.  The range of spike-in concentrations used in Dataset II is shifted downwards from that of datasets I and III, and as such may not be adequately probing the saturation region to give acceptable fits in all cases.  Furthermore, the spike-in concentrations at the lower end may be probing the regime in which the target concentration is depleted during the hybridisation step, which is beyond the applicability of the model leading to Eq.~\ref{hyperbolicResponse} which we describe below.  The analyses in Sections~\ref{sec:Qualitative} and \ref{sec:Quantitative} below are restricted to the subset of fits to Eq.~\ref{hyperbolicResponse} for which all three parameters $A$, $B$ and $K$ are positive.  Table~\ref{tab:coef_of_var} gives the coefficient of variation of the fitted data for each dataset, and the percentage of features for which an acceptable fit with three positive parameters to the hyperbolic response function was obtained.  In general, agreement with a hyperbolic response curve with positive parameters is excellent for datasets I and III, and reasonable for dataset II. 

\begin{table}
\caption{\small Coefficient of variation of the data, assumed to distributed from a Gamma distribution with constant coefficient of variation within any one dataset about a mean given by Eq.~\ref{hyperbolicResponse}.  The two right hand columns give the percentage of probesets which for which the fit gives positive values to all three parameters $A$, $B$ and $K$.}
\label{tab:coef_of_var}
\begin{center}
\begin{tabular}{|l|c|c|c|}
\hline
Dataset & Coef. of & \multicolumn{2}{r} \% of accepted fits \\
& variation & PM & MM \\
\hline
  I      &  0.12  &  97.9  &  91.6 \\
  II     &  0.14  &  72.5  &  37.5 \\
  III    &  0.17  &  100   &  98.4 \\
\hline
\end{tabular}
\end{center}
\end{table}

\section{The model}
\label{sec:Model}

Consider the response of a given feature to a spike-in concentration $x$ of a particular RNA transcript in the presence of an unknown complex background of non-specific target RNA.  We write the measured fluorescence intensity $I(x)$ in the form 
\begin{equation}
I(x) = a + b \theta(x), \label{intensity}
\end{equation}
where $a$ is a physical background due to effects unrelated to fluorescent label carrying duplexes, such as reflection from the glass surface of the microarray, and $b$ is the maximum fluorescence intensity, that is, the contribution from fluorescent dye if all probes on the feature were occupied with labelled probe-target duplexes.  It is argued in ref.~\cite{Binder06a} that the maximum intensity $b$ should vary only weakly due to differing probe sequences.  Throughout this paper we assume $a$ and $b$ to be fixed constants for a given microarray.  The coverage fraction, $\theta(x)$, is the fraction of probes on the feature carrying probe-target duplexes at the time of scanning.  It satisfies $0 \le \theta(x) \le 1$. 

The coverage fraction is determined by a number of reactions between various chemical species.  The species and reactions considered in our model are set out in Tables~\ref{tab:species} and \ref{tab:reactions} respectively.  The first five reactions in Table~\ref{tab:reactions} are assumed to reach equilibrium during the hybridisation step. The rate constants $K_i^\bulk, K^\Sfold, K^\spec, K_i^\NS$ and $K^\Pfold$ are the ratio of the forward to backward rates for each reaction.  The washing phase, which is primarily designed to remove unbound targets before scanning, also dissociates some duplexes\cite{Burden06,Held06}.  Thus the last two reactions are unidirectional as dissociated duplexes are continuously flushed out of the cartridge and replaced with a buffer solution containing no RNA.  

\begin{table}
\caption{\small Chemical species present in the model.}
\label{tab:species}
\begin{center}
\begin{tabular}{|l|l|}
\hline
$S$     &  Single strand PM-feature-specific RNA target in solution \\
$NS_i$    &  Single strand non-specific RNA target in solution of species $i$ \\
$S.NS_i$ & Bound RNA/RNA duplex in solution unavailable for hybridisation \\
$S'$     & Folded specific target in solution rendered unavailable for hybridisation \\
$P$     &  Unbound probe at the microarray surface \\
$P.S$  & Duplex formed from PM-feature-specific RNA target binding to probe \\
$P.NS_i$  & Duplex formed from non-specific RNA target binding to probe \\
$P'$     & Folded probe at the microarray surface rendered unavailable for hybridisation \\
\hline
\end{tabular}
\end{center}
\end{table}

\begin{table}
\caption{\small Chemical reactions occurring in the model.  Rate constants, defined as the ratio of forward to backward reaction rates, are given in the right hand column.}
\label{tab:reactions}
\begin{center}
\begin{tabular}{|l|l|c|l|}
\hline
In bulk solution & Non-specific hybridisation & $S + NS_i \rightleftharpoons S.NS_i$ & $K_i^\bulk$ \\
                            & Specific target folding & $S \rightleftharpoons S'$ & $K^\Sfold$ \\
\hline
At the microarray & Specific hybridisation  & $P + S \rightleftharpoons P.S$ & $K^\spec$ \\
surface & Non-specific hybridisation & $P + NS_i \rightleftharpoons P.NS_i$ & $K_i^\NS$ \\
 & Probe folding & $P \rightleftharpoons P'$ & $K^\Pfold$ \\
\hline
During the & Dissociation of specific duplexes & $P.S \rightarrow P\, (+ S)$ & \\
 washing phase & Dissociation of non-specfic duplexes & $P.NS_i \rightarrow P\, (+ NS_i)$ & \\
\hline
\end{tabular}
\end{center}
\end{table}

The effect of the first two reactions, bulk hybridisation and specific target folding in solution, is to reduce the concentration of specific target available for hybridisation onto the microarray from its spike-in value of $x$ to a value $[S]$, that is the concentration of single strand RNA target $S$ in solution.  (Following the usual convention, square brackets indicate concentrations.)  For these reactions we follow the analysis of ref.~\cite{Heim06}.  The label $i$ in Table~\ref{tab:reactions} ranges over all possible subsequences of the 25-mer part of the specific target RNA sequence complementary to the PM probe.  The species $NS_i$ in this reaction includes any target RNA molecule containing a subsequence complementary to the $i$th subsequence.  At equilibrium, we have
\begin{equation}
\frac{[S']}{[S]} = K^\Sfold, \qquad \frac{[S.NS_i]}{[S][NS_i]} = K_i^\bulk.    
\end{equation}
The relation $x = [S] + [S'] + \sum_i [S.NS_i]$ then gives 
\begin{equation}
[S] = \frac{x}{1 + K^\Sfold + X^\bulk},      \label{Sformula}
\end{equation}
where
\begin{equation}
X^\bulk = \sum_i [NS_i] K_i^\bulk.   \label{XbulkDef}
\end{equation}

The next three reactions, occurring at the microarray surface, determine the duplex coverage fraction of the feature at the end of the hybridisation step, and before washing.  Let the fraction of probes on the feature that have formed duplexes with either specific or non-specific target mRNA molecules and survived to a time $t_W$ after the commencement of the washing process be $\theta(x, t_W)$.  We split the fraction of probes which have formed duplexes at the end of the hybridisation step and before washing into two contributions: 
\begin{equation}
\theta(x,0) = \theta^\spec + \theta^\NS. \
\end{equation}
The first contribution, $\theta^\spec \propto [P.S]$, is that due to duplexes formed with specific mRNA targets, and the second contribution, $\theta^\NS \propto \sum_i[P.NS_i]$, is that due to duplexes which have formed with non-specific mRNA targets, the sum being over targets containing a subsequence complimentary to the $i$th subsequence of the probe.  

Either by balancing equilibrium concentrations against chemical rate constants~\cite{Binder04} or by considering the Gibbs distribution of states at constant chemical potential~\cite{Halperin04,Burden06} one obtains the isotherms 
\begin{eqnarray}
\theta^\spec & = & \frac{X^\spec}{1 + K^\Pfold + X^\spec + X^\NS}     \label{thetaS} \\
\theta^\NS & = & \frac{X^\NS}{1 + K^\Pfold + X^\spec + X^\NS}     \label{thetaNS}
\end{eqnarray}
where, following the notation of ref.\cite{Binder06a}, we define
\begin{equation}
X^\spec = [S] K^\spec, \qquad X^\NS = \sum_i [NS_i] K_i^\NS.    \label{XsurfaceDef}
\end{equation}
The calculation leading to these isotherms assumes negligible depletion of target molecules in bulk solution during the hybridisation process.  

Note that, in the asymptotic limit of high spike-in concentration, namely $x \rightarrow \infty$ while holding $[NS_i]$ constant, Eqs.~\ref{Sformula} to \ref{XsurfaceDef} imply $\theta^\spec \rightarrow 1$ and  $\theta^\NS \rightarrow 0$, implying that the feature becomes saturated with specific duplexes. This is contrary to the differing isotherm asymptotes observed empirically.  To explain the differing asymptotes, we include in our model the final two reactions in Table~\ref{tab:reactions}, namely the washing step~\cite{Burden06}.  Define the probability that a given probe-target duplex has survived up to a washing time $t_W$ to be $s^\spec(t_W)$ for a specific duplex and $s_i^\NS(t_W)$ for a non-specific duplex of species $i$.  The survival functions $s^\spec$ and $s_i^\NS$ depend on probe and target base sequences, satisfy $s^\spec(0) = s_i^\NS(0) = 1$, are positive and are monotonically decreasing.  Defining an average non-specific survival function $s^\NS(t_W)$ by 
\begin{equation}
X^\NS s^\NS(t_W) = \sum_i [NS_i] K_i^\NS s_i^\NS(t_W),   \label{NSsurvival}
\end{equation}
the coverage fraction at washing time $t_W$ is then 
\begin{equation}
\theta(x, t_W) = \theta^\spec s^\spec(t_W) + \theta^\NS s^\NS(t_W).
\end{equation}
Substituting Eqs.~\ref{thetaS} and \ref{thetaNS} and rearranging gives 
\begin{eqnarray}
\theta(x, t_W) & = &  \frac{X^\NS s^\NS(t_W)}{1 + K^\Pfold + X^\NS} +   \\
& & 
      \left(s^\spec(t_W) - \frac{X^\NS s^\NS(t_W)}{1 + K^\Pfold + X^\NS}\right) 
         \frac{(1 + K^\Pfold + X^\NS)^{-1} X^\spec}{1 + (1 + K^\Pfold + X^\NS)^{-1} X^\spec}.  \nonumber
\end{eqnarray}
Finally, with help from Eqs.~(\ref{Sformula}) and (\ref{XsurfaceDef}), and suppressing the $t_W$ dependence, we get
\begin{equation}
\theta(x) = \alpha + \beta\frac{Kx}{1 + Kx},    \label{thetaIsotherm}
\end{equation}
where
\begin{eqnarray}
\alpha & = & \frac{X^\NS s^\NS}{1 + K^\Pfold + X^\NS},  \label{alphaModel} \\
\beta    & = & s^\spec - \alpha,                                               \label{betaModel}  \\
K          & = &  \frac{K^\spec}{(1 + K^\Sfold + X^\bulk)(1 + K^\Pfold + X^\NS)} .   \label{KModel}
\end{eqnarray}

This model, summarised by Eqs.~(\ref{intensity}) and (\ref{thetaIsotherm}) is consistent with the empirical observation of Eq.~(\ref{hyperbolicResponse}), with
\begin{equation}
A = a + b\alpha$, \qquad $B = b\beta.   \label{ABab} 
\end{equation}  
Eqs.~\ref{alphaModel} to \ref{ABab} (with the help of Eqs.~\ref{XbulkDef}, \ref{XsurfaceDef} and \ref{NSsurvival}) relate the empirically fitted parameters $A$, $B$ and $K$ to underlying physical quantities, namely  $a$, $b$, the concentrations of chemical species in Table~\ref{tab:species}, reaction rates in Table~\ref{tab:reactions} and washing survival functions $s^\spec$ and $s^\NS_i$.   

Physical effects which have not been included in the model include target depletion, which should only manifest at extremely low target concentrations, incomplete probe synthesis during the manufacturing process~\cite{Forman98}, and probe-probe interactions.  Each of these effects will, in theory, cause the response curve to deviate from a hyperbolic form.   A discussion of probe-probe interactions, for instance, can be found in ref.~\cite{Burden06}.  The choice of model in this paper is guided by a desire to balance complexity of the problem with practicality.  

\section{Qualitative behaviour of the fits}
\label{sec:Qualitative}

\begin{figure}
\includegraphics[scale=0.55]{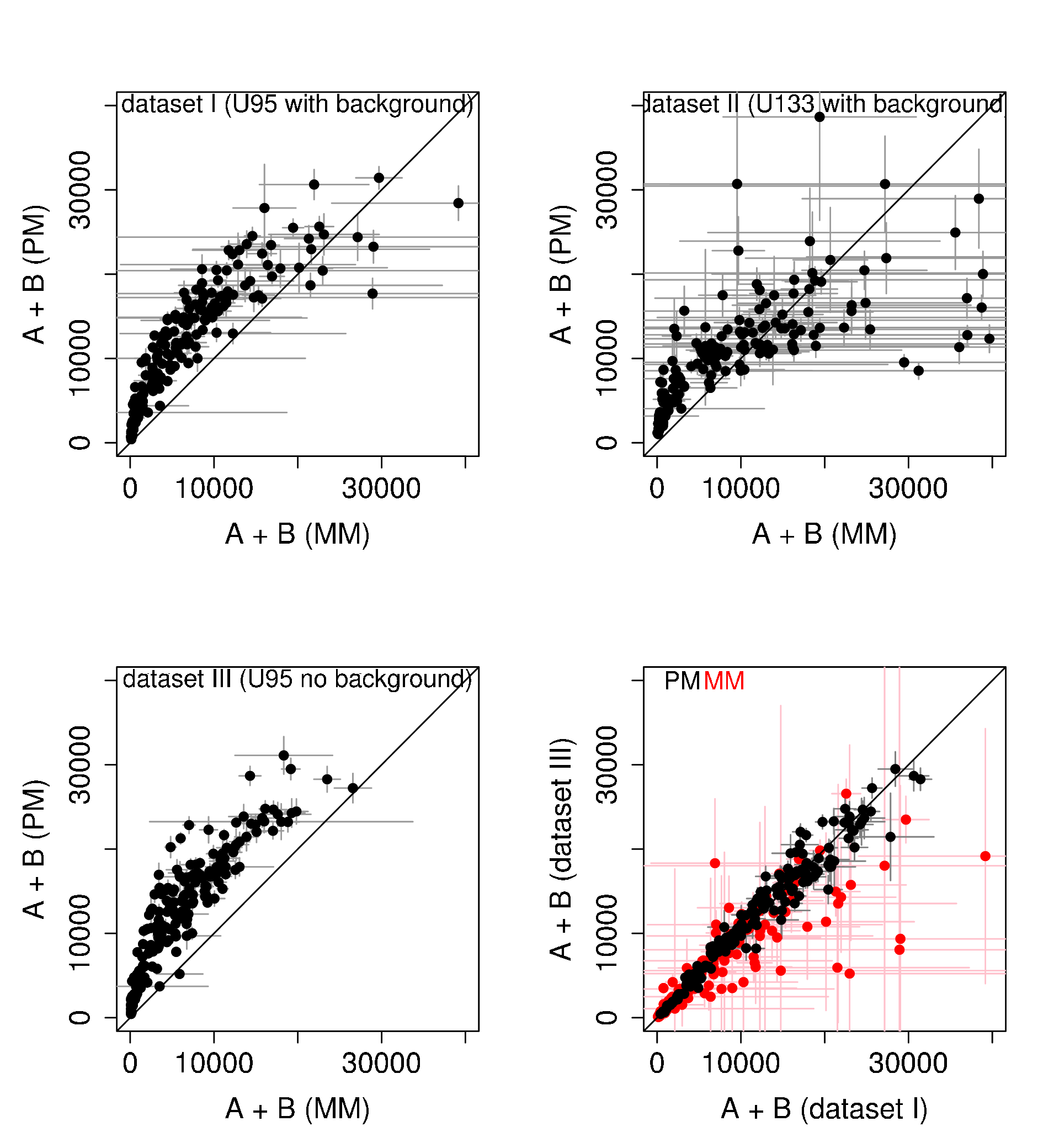}
\centering
\caption{The first three panels show fitted asymptotes $I(\infty) = A + B$, defined in Eq.~\ref{hyperbolicResponse} for PM/MM pairs for each of the three datasets.  The fourth panel compares  the asymptotes for dataset I (with non-specific background) with those for dataset III (without non-specfic background).  Standard errors arising from the fits to Eq.~\ref{hyperbolicResponse} are also shown.  }
\label{fig:AplusB_comparison}
\end{figure}

Before considering a detailed analysis of the ability of the model to explain the parameters of the empirical fits, one can carry out a number of simple qualitative checks.   The first three panels of Figure~\ref{fig:AplusB_comparison} compare the fitted saturation asymptotes $A + B$ for PM/MM pairs of features for each of the three datasets.  Consistent with the hypothesis that a portion the bound probe-target duplexes are removed during the washing step, the asymptotes cover a broad range of values.  The MM asymptote is almost always less than its PM partner, consistent with the scenario that a saturated feature of PM duplexes will lose less duplexes to washing than the partner saturated feature of less tightly bound MM duplexes~\cite{Burden06}.  The observed pattern breaks down at higher values of $A + B$, as the fits must be extrapolated further past the highest spike-in concentration to estimate the asymptote, and numerical accuracy is lost.  This is most evident for dataset II, for which spike-in concentrations only extend to 512\,pM, compared with 1024\,pM for datasets I and III.  

\begin{figure}
\includegraphics[scale=0.6]{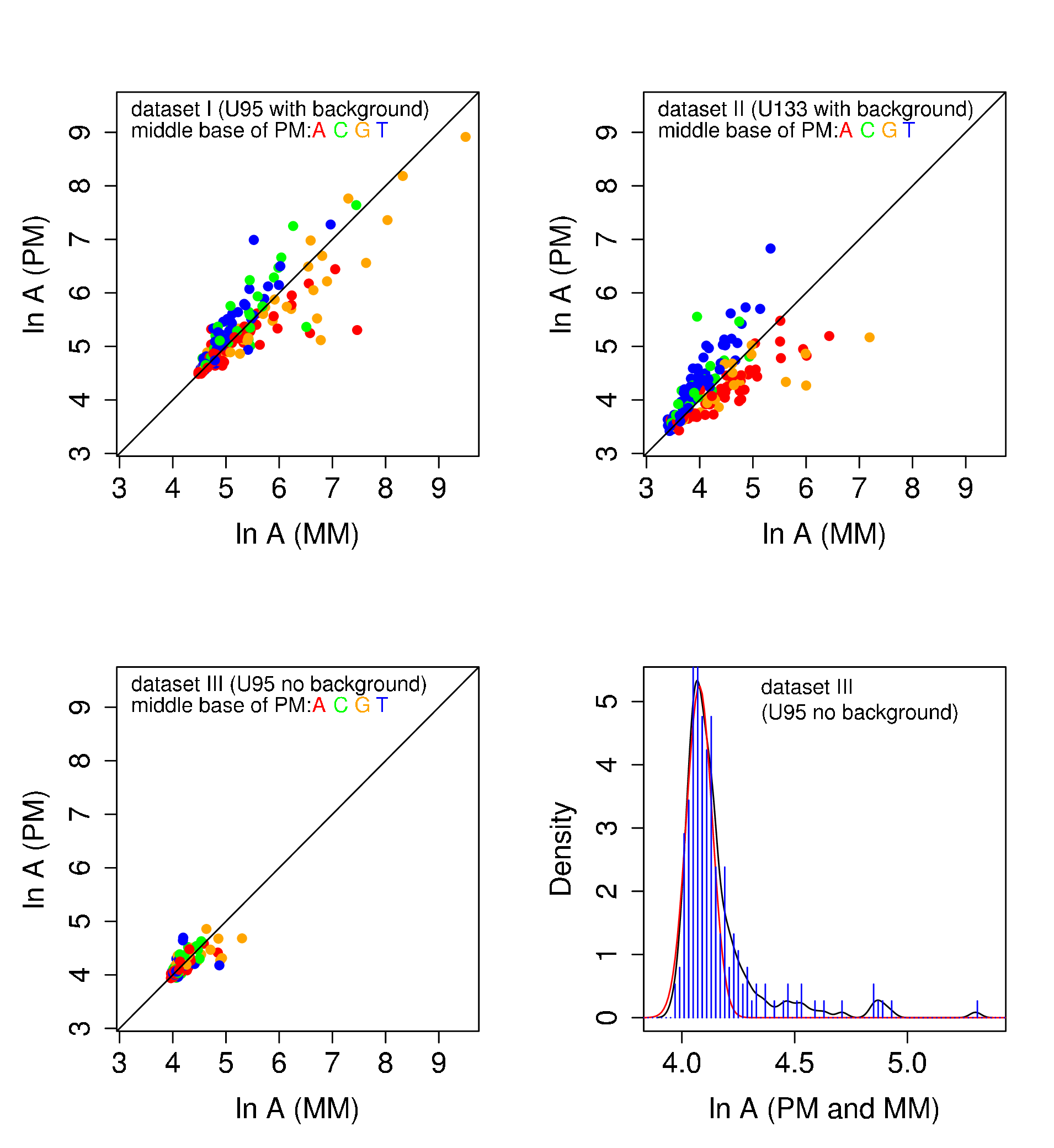}
\centering
\caption{The log of the fitted baseline fluorescence intensities $I(0) = A$, defined in Eq.~\ref{hyperbolicResponse}, for PM/MM pairs for datasets I and III.  The middle (13th) base of the PM probe sequence is colour coded as indicated.  The fourth panel shows a histogram of $\ln A$ (in blue), a kernel density estimate of the histogram using a gaussian kernel with a bandwidth of 0.025 (in black), and a fit of the left part of the histogram to a Gamma distribution in $A$ with a mean of 59 and a coefficient of variation of 0.057 (in red).}
\label{fig:A_comparison}
\end{figure}

From Eqs.~\ref{betaModel} and \ref{ABab}, the saturation asymptote is given by $I(\infty) = A + B = a + bs^\spec$.  This depends only on the rate $s^\spec$ at which specific duplexes are dissociated by the washing process, and not on the properties of non-specific duplexes.  Thus the model predicts that the asymptote of the response function is unaffected by non-specific hybridisation.  The fourth panel of Figure~\ref{fig:AplusB_comparison}, which compares the asmptote for dataset I, (U95 with a complex non-specific background), with that for dataset III, (U95 without a non-specific background), confirms this.    That is, the asymptote $I(\infty)$ for any feature is the same for dataset I as for dataset III to within the standard errors of the isotherm fits.

The parameter $A$ in Eq.~\ref{hyperbolicResponse} is the baseline intensity estimate at zero spike-in concentration.  From Eq.~\ref{ABab}, it consists of a physical background component $a$, and a component due to non-specific hybridisation, $b\alpha$.  Consistent with this, the $A$ values, shown in Fig.~\ref{fig:A_comparison}, are spread over a much broader range for datasets I and II in which a complex mRNA background was present than for dataset III with no background and therefore little non-specific hybridisation.  

An obvious pattern, which emerges from comparing $A$ from PM/MM pairs of features in datasets I and II, is that non-specific hybridisation is stronger for a probe whose middle base is a pyrimidine (C or T) than for its partner probe whose middle base is a purine (A or G).  This effect has been noted previously for microarray intensity data generally, and there is some debate about the its physical origins~\cite{Naef03,Binder05,Carlon06b}.  The effect is better understood at the level of individual letter frequencies.  Binder et al.\cite{Binder04} have noted that probe sensitivities increase with C-content, and decrease with A-content, while the G- or T-content of the probe has little effect.  There are probably two effects contributing to this pattern.  Firstly, the averaged contributions to DNA/RNA binding energies calculated from nearest neighbour stacking models~\cite{Sugimoto95} are ordered as~\cite{Binder04b,Carlon06b} 
\begin{equation}
|\Delta G^{\rm av}_C| > |\Delta G^{\rm av}_G| \approx |\Delta G^{\rm av}_T| > |\Delta G^{\rm av}_A|,
						\label{DeltaGorder}
\end{equation}
causing the substitution of a pyrimidine by a purine to decrease probe sensitivity and vice versa.  Secondly, there is the simple geometric effect that pyrimidines, having a small single ring structure, will tolerate mismatches more easily than purines, which have a double ring structure, and would therefore need to deform the molecular backbone to bind with a target closely matching the probe sequence either side of the central base.  No obvious pyrimidine/purine asymmetry is observed in the $A$ values from dataset III.  This is to be expected as the parameter $A$ in this case is essentially the physical background without hybridisation contributions. 

The fourth panel of Fig.~\ref{fig:A_comparison} shows a histogram of $A$ values from dataset III, for which there is no complex background present.  There is very little non-specific hybridisation, and most of this data represents statistical noise in the physical background parameter $a$ (defined in Eq.~\ref{intensity}).  Ignoring the tail, which we assume to be due to a small amount of non-specific hybridisation from the other spiked-in transcripts in the latin square protocol, we estimate $a$ by fitting the left hand part of the histogram to a gamma distribution in the unlogged data.  The fitted distribution has a mean of 59 and a coefficient of variation of 0.057.  

\begin{figure}
\includegraphics[scale=0.6]{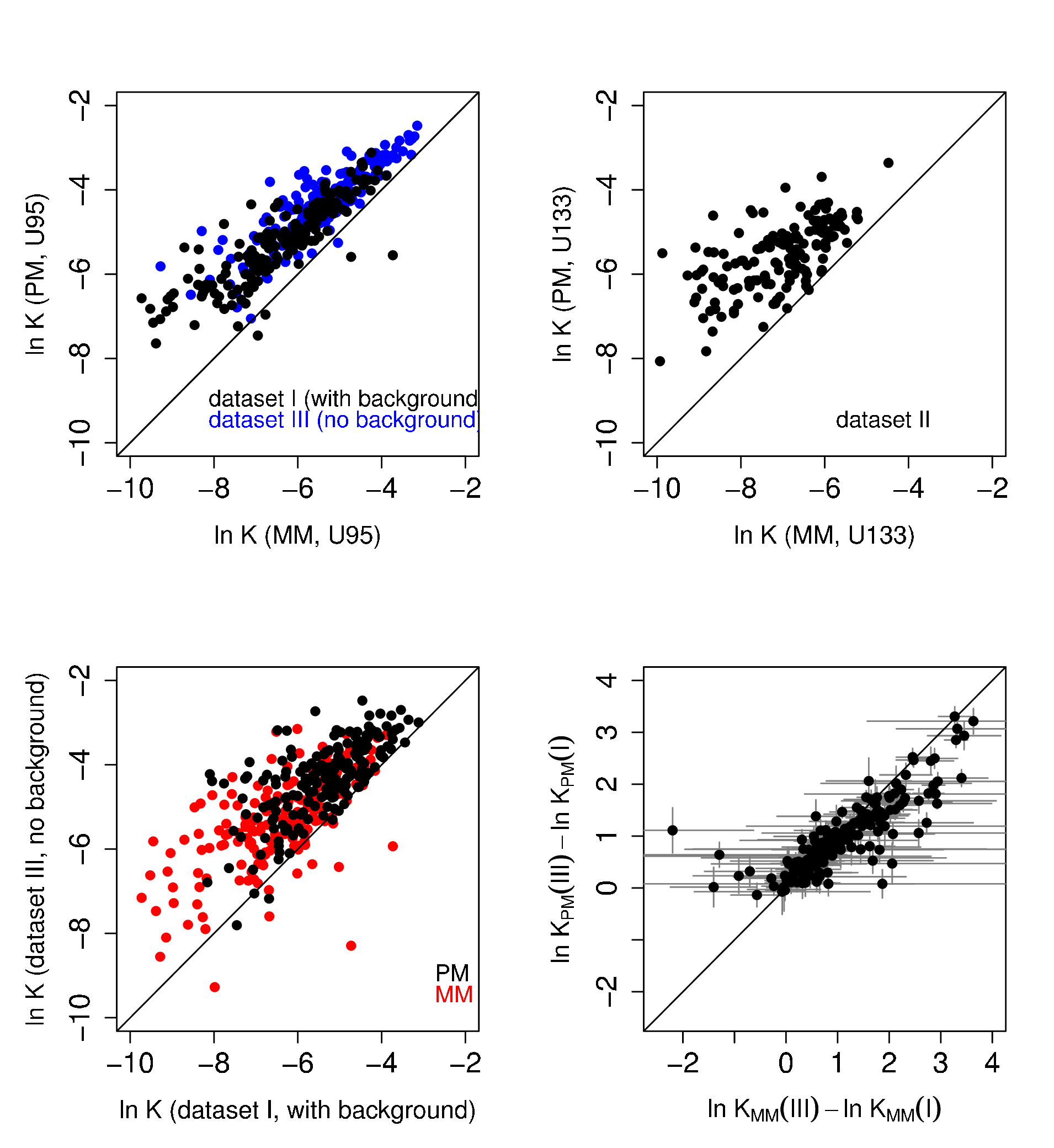}
\centering
\caption{The log of the fitted parameters $K$, defined in Eq.~\ref{hyperbolicResponse}.  The first and second panels compare PM/MM pairs for each of the three datasets.  The third panel compares datasets I and III, with and without the complex human pancreas  background respectively.  The fourth panel compares the increase in $\ln K$ for MM with that for PM when the complex human pancreas background is removed.}
\label{fig:K_comparison}
\end{figure}

Various comparisons of the effective adsorption rate constant $K$ are shown in Fig.~\ref{fig:K_comparison}.  This parameter is modelled in terms of more fundamental rate constants via Eq.~\ref{KModel}.  One can reasonably assume the differences between PM and MM for the indirect effective rate constants $X^\bulk$, $K^\Pfold$ and $X^\NS$ to be small compared with that for the specific rate constant $K^\spec$, because of averaging over large numbers of non-specific species.  Thus, to a reasonable approximation, $\ln K_\PeM/K_\MM \approx \ln K_\PeM^\spec/K_\MM^\spec \approx -\Delta\Delta G/RT$, where $\Delta\Delta G$ is the difference in specific binding energies between a specific mRNA target and a PM and MM probe respectively.  This empirical result has been noted previously as a shift away from the diagonal in a plot of $K_\PeM$ versus $K_\MM$ for dataset I~\cite{Hekstra03} and for dataset II~\cite{Binder06b}, and is confirmed here for all three datasets.  

Estimates of $-\Delta\Delta G$ were obtained by taking a weighted average of $-\Delta\Delta G_p = (\ln K_\PeM - \ln K_\MM)_p$ over fitted isotherms $p$ for each of the central base letters.  The weighting $\sum_p \left(-\Delta\Delta G_p/ s_p^2\right)/\sum_p\left(1/ s_p^2\right)$, where $s_p$ is the standard error in $\Delta\Delta G_p$ estimated from the standard error in $K_p$ arising from the isotherm fit is chosen to minimise the error in the average.  The results, given in Table~\ref{tab:DeltaGfromK}, are consistent with the ordering of binding energies calculated from nearest neighbour stacking models in Eq.~\ref{DeltaGorder}.

\begin{table}
\caption{\small $-\Delta\Delta G$ estimated from a weighted average of ($\ln K_\PeM - \ln K_\MM$) for each of the four central bases.  }
\label{tab:DeltaGfromK}
\begin{center}
\begin{tabular}{lllll}
\hline
 & C & G & T & A \\
\hline
Dataset I   &   $1.09\pm0.05$ & $0.94\pm0.06$ & $0.98\pm0.04$ & $0.74\pm0.04$  \\
Dataset II  &   $1.74\pm0.14$ & $1.43\pm0.10$ & $1.26\pm0.05$ & $0.84\pm0.06$  \\
Dataset III &   $1.10\pm0.04$ & $0.90\pm0.04$ & $0.97\pm0.03$ & $0.66\pm0.03$  \\
\hline
\end{tabular}
\end{center}
\end{table}

The third panel of Fig.~\ref{fig:K_comparison} compares the effective rate constant $K$ for the U95 experiments with and without the complex human pancreas background.  From Eq.~\ref{KModel}, one sees that removing the background, which has the effect of setting $X^\bulk$ and $X^\NS$ to zero, should increase $K$.  This is confirmed in the plot, and is also evident as a shift in the inflection points to the left between Figs.~\ref{fig:fit37777} and \ref{fig:fit37777nbg}.  The fourth panel compares the amount by which $\ln K$ increases as the complex background is removed for PM and MM.  We observe that removing the effects of $X^\bulk$ and $X^\NS$ affects $K$ for a PM probe and its MM partner by a similar factor.  The small number of points away from the diagonal line to the left of the plot are caused by the difficulty in fitting the MM isotherm when $K_\MM^{-1}$ is beyond the upper limit of the range of spike-in concentrations.  

\section{Quantitative behaviour of the fits}
\label{sec:Quantitative}

In this section we explore the ability of the model to explain the quantitative relationship between the fitting parameters of the hyperbolic response curve Eq.~\ref{hyperbolicResponse} and known physical properties of microarrays.  We divide the analysis into two parts: The parameters $A$ and $B$ which set the vertical scale of the hyperbolic response curve, and the parameter $K$ which sets its horizontal scale.  

\subsection{Vertical scale parameters}
\label{sec:Vertical}

The parameters $A$ and $B$ are explained in the model in terms of the more fundamental quantities $a$, the physical background and $b$, saturation intensity above background (which together set the intensity scale in terms of the dimensionless duplex coverage fraction) and $\alpha$ and $\beta$ (which are driven by the chemical reactions in Table~\ref{tab:reactions}).  

\begin{figure}
\includegraphics[scale=0.6]{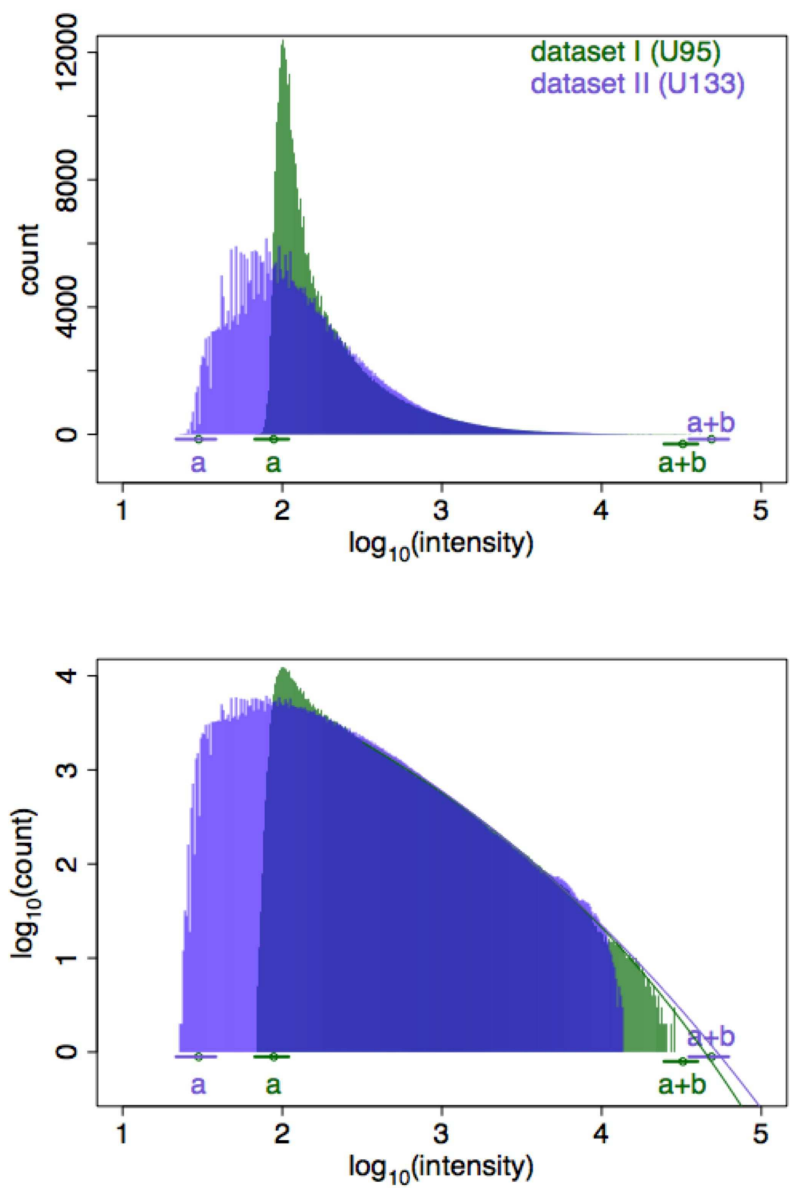}
\centering
\caption{Histograms of quantile normalised fluorescence intensities across microarrays in datasets I and II on a linear (upper) and logarithmic (lower) scale.  Both PM and MM intensities are included.  Counts are from bins of size 0.01 on the log intensity axis.  Also indicated are estimates of the parameters $a$ and $b$ for each dataset, with bars indicating two standard deviations of the spread in the intensity data either side. The curves fitted to the histograms in the lower panel are explained in Section~\ref{sec:Prediction}.}
\label{fig:histogram}
\end{figure}

We begin with $a$ and $b$, which are assumed to be fixed for an entire microarray.  In Fig.~\ref{fig:histogram} are plotted histograms of measured fluorescence intensities across microarrays from datasets I and II.  Because these data have been quantile normalised, a common histogram will apply to all microarrays within a given dataset.  In the absence of statistical noise, the parameters $a$ and $a + b$ should provide bounding limits for these histograms.  However, given the coefficients of variation reported in Table~\ref{tab:coef_of_var}, the raw intensity measurements could well extend beyond these limits.  According to Eq.~\ref{ABab} the parameter $a$ should also be a lower cutoff on the fitted parameter $A$ (corresponding to the case of negligible non-specific hybridisation), while the analysis of $A$ in Section~\ref{sec:Qualitative} for dataset III (see the fourth panel of Fig.~\ref{fig:A_comparison}) suggests a much smaller coefficient of variation for the fitted value of $A$ than for the intensity data generally.  For these reasons we use as an estimate of $a$ the minimum over all fits within a dataset of the parameter $A$.  These estimates are shown in Fig.~\ref{fig:histogram}, together with bars extending two standard deviations either side using the coefficients of variation in Table~\ref{tab:coef_of_var}.  

\begin{table}
\caption{\small Fitted parameters, to 3 significant figures, occuring in the analysis of Section~\ref{sec:Quantitative}.  }
\label{tab:parameters}
\begin{center}
\begin{tabular}{|c|c|r|r|r|}
\hline
Defining equation & Parameter & Dataset I & Dataset II & Dataset III\\
\hline
 (\ref{intensity}) & $a$  &  88.4 &  30.0 & \\
                  &  $\log_{10} a$ &  1.95   &  1.48 &   \\
                 & $b$   & 32300 & 48800 &  \\
                 &  $\log_{10} b$  &  4.51   &  4.69  &  \\ 
\hline
  (\ref{kappa_def})    &   $c_0$              &  62.2 & 37.9 &  \\
                                    &   $\lambda_0$ & 0.0920  & 0.0841 & \\
\hline
 (\ref{alpha_m2})      &   $c_1$          & $-16.1$ & $-14.8$ & \\
 ($\alpha$ Model 2) &   $c_2$          & $-0.186$ & $-0.148$ & \\
                                    &   $c_3$          & $0.124$ & $0.102$ & \\
\hline
 (\ref{alpha_m5})      &   $c_1$          & $-14.8$ & $-14.8$ & \\
 ($\alpha$ Model 5) &   $c_2$          & $-0.200$ & $-0.149$ & \\
                                    &   $c_3$          & $0.0776$ & $0.101$ & \\
                                    & $\lambda_\alpha$ & 0.176 & 0.909 & \\
                                    & $\mu_\alpha$         & 4.57 & -6.14 & \\
\hline
 (\ref{model_K7})   &   $\lambda_\spec$ &  0.145  &  0.0824  & 0.0944  \\
  ($K$ Model 7)     &   $         \mu_\spec$ &  $-62.9$    &  69.4  &  $-73.4$  \\
                                 &   $\lambda_\Sfold$ &   0.204 &  0.131  &   0.202 \\
                                 &   $         \mu_\Sfold$ &  $-59.3$ &  $-51.5$  &  $-73.7$  \\
                                 &   $\lambda_\Pfold $&  0.268  &  0.100  &   0.385 \\
                                 &   $         \mu_\Pfold$ &  $-0.757$  &  136  &  0.917  \\
\hline
\end{tabular}
\end{center}
\end{table}

\begin{figure}
\includegraphics[scale=0.55]{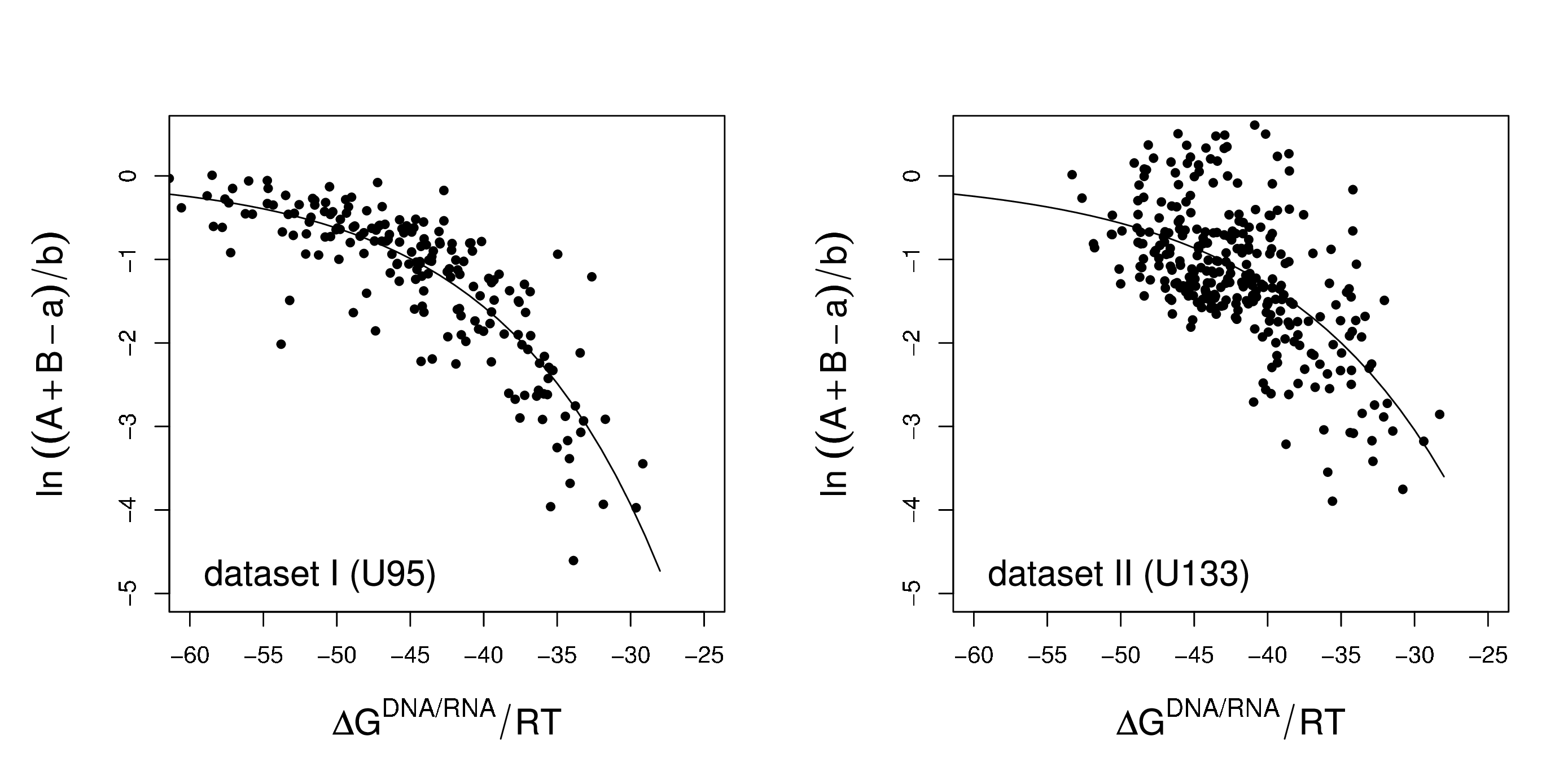}
\centering
\caption{Fits of Eq.~\ref{AplusB_fit} to the parameter combination $A + B$ of the hyperbolic response function fits to the PM data for datasets I and II. The parameter $b$ has been absorbed into a shift in the ordinate.}
\label{fig:AplusB_fit}
\end{figure}

To estimate the saturation parameter $b$ from fits to the hyperbolic response function, and to explain the observed values of the combination $\alpha + \beta$, we make use of the model's prediction that  the asymptotic intensity at high spike-in concentration, $I(\infty)$, is  determined by the washing-step survival function of PM-specific duplexes, $s^\spec(t_W)$~\cite{Burden06}.  Thus from Eqs.~\ref{betaModel} and \ref{ABab} we have 
\begin{equation}
I(\infty) = A + B = a + b(\alpha + \beta) = a + b s^\spec(t_W) = a + b e^{-\kappa t_W}, 
\end{equation}
where we assume a uniform washing rate $\kappa$ that depends only on the probe and target nucleotide sequences via their binding affinity.  Following Ref.~\cite{Burden06}, we model $\kappa$ in terms of the RNA/DNA duplex free binding energy in bulk solution: 
\begin{equation}
\kappa t_W =  c_0 e^{\lambda_0 \Delta G^{\rm DNA/RNA} /(RT)}.         \label{kappa_def}
\end{equation}
Here $\Delta G^{\rm DNA/RNA}$ is calculated using the nearest neighbour stacking model and parameters of Ref.~\cite{Sugimoto95}, $R$ is the gas constant, $T$ the absolute temperature and $c_0$ and $\lambda_0$ are undetermined constants.  We use the convention that $\Delta G^{\rm DNA/RNA}$ is negative for a bound state.  Rearranging gives 
\begin{equation}
\ln(\alpha + \beta) = \ln\frac{A + B - a}{b} = - c_0 e^{\lambda_0 \Delta G^{\rm DNA/RNA} /(RT)},  \label{AplusB_fit}
\end{equation} 
where the $A$ and $B$ are determined for each feature from fitted hyperbolic response functions, $a$ has been estimated above, and $\ln b$, $c_0$ and $\lambda_0$ are fitting parameters.  Fits to datasets I and II are shown in Fig.~\ref{fig:AplusB_fit}, and fitting parameters listed in Table~\ref{tab:parameters}.  The fits were done by minimising the sum of the squares of the residuals with respect to the three fitting parameters.  

The parameter $\alpha$ is in principle predicted by the model in terms of fundamental physical constants via Eqs.~\ref{NSsurvival} and \ref{alphaModel}.  It is determined mainly by non-specific hybridisation, including a strong influence from the probe sequence pyrimidine content as suggested by Fig.~\ref{fig:A_comparison}, and by probe folding.  Without a detailed knowledge of the composition of the non-specific target solution, a direct evaluation of $\alpha$ is of course impossible.  However, one can aim for an ansatz in terms of those quantities which are known.   Following the reasoning used above, the washing-step survival function of the $i$th non-specific species $s_i^\NS(t_W)$ can be assumed to take the form of a double exponential function of the corresponding free binding energy $\Delta G^{\rm DNA/RNA}_i$.  The value of the double exponential function ($f(x)  = e^{-e^{x}}$) undergoes a changeover from 1 for $x<<0$ to 0 for $x>>0$ over a narrow range of its argument.   Thus the  factor $s_i^\NS(t_W)$ in Eq.~\ref{NSsurvival} can be thought of as a switch with removes from the sum any binding configuration $i$ less tightly bound than some threshold.  

The numerator of Eq.~\ref{alphaModel} is then a sum of reaction rate constants $K_i^{\NS}$, weighted by the number of target molecules in solution with nucleotide sequences complementary to the $i$th subsequence of the probe.  Numerical estimates carried out in the context of bulk hybridisation in solution\cite{Heim06} show that such a sum can be well approximated as an exponential function of free binding energy of the entire probe sequence to its complement.  Assuming then that the behaviour of $\alpha$ is dominantly exponential in $\Delta G^{\rm DNA/RNA}$, and taking into account the added effect of the probe's pyrimidine count, we have tested the following nested models: 
\begin{eqnarray}
\mbox{Model~0:  } & & \ln \alpha  =  c_1  + \epsilon, \label{alpha_m0} \\
\mbox{Model~1:  } & & \ln \alpha  =  c_1 + c_2 \Delta g^{\rm DNA/RNA} + \epsilon, \label{alpha_m1} \\
\mbox{Model~2:  } & & \ln \alpha =  c_1 + c_2 \Delta g^{\rm DNA/RNA} + c_3 n_{\rm pyr}  + \epsilon,  \label{alpha_m2} \\
\mbox{Model~3:  } & & \ln \alpha =   c_1 + c_2 \Delta g^{\rm DNA/RNA} + c_3 n_{\rm Pyr} + c_4 n_{\rm Pyr} \Delta g^{\rm DNA/RNA} + \epsilon,  \label{alpha_m3}
\end{eqnarray}
where $c_1, \ldots, c_4$ are fitting parameters to be determined, $\Delta g^{\rm DNA/RNA} = \Delta G^{\rm DNA/RNA}/(RT)$, $n_{\rm pyr}$ is a count of the number of pyrimidines in the 25-mer probe sequence and $\epsilon$ is the residual error, which is assumed Gaussian.   

To include the effect of probe folding, we approximate $K^\Pfold$ in Eq.~\ref{alphaModel} by a single exponential term 
\begin{equation}
K^\Pfold  =  \exp[\lambda_\alpha(\mu_\alpha - \Delta g^{\rm DNA-fold})],  
\end{equation}
where $\lambda_\alpha$ and $\mu_\alpha$ are fitting parameters and $\Delta g^{\rm DNA-fold} = \Delta G^{\rm DNA-fold}/(RT)$, where $\Delta G^{\rm DNA-fold}$ is calculated for each 25-mer probe sequence from Zuker's Mfold web server~\cite{Zuker03} with the temperature set to $45^\circ$C and other parameters set to their default values.  The Mfold web server calculates the free energy of the most energetic folding configuration of a given single strand DNA sequence, though ideally one should include a Boltzman weighted sum over all possible folding configurations.  Models~1 and 2 are then nested within Models~4 and 5 respectively:
\begin{eqnarray}
\mbox{Model~4:  } & & \ln \alpha  =  c_1 + c_2 \Delta g^{\rm DNA/RNA}      \nonumber\\
                      & &  - \ln\{1 + \exp[\lambda_\alpha(\mu_\alpha - \Delta g^{\rm DNA-fold})]\}  + \epsilon,    \label{alpha_m4}  \\
\mbox{Model~5:  } & & \ln \alpha =  c_1 + c_2 \Delta g^{\rm DNA/RNA} + c_3 n_{\rm pyr}  \nonumber\\
                      & &  - \ln\{1 + \exp[\lambda_\alpha(\mu_\alpha - \Delta g^{\rm DNA-fold})]\}  + \epsilon.      \label{alpha_m5} 
\end{eqnarray}

The above nested models can be tested for the significance of the extra parameters introduced in going from a less to a more complicated model.  For instance, to test the significance of the extra parameter distinguishing model $m_2$ from the simpler  $m_1$, consider for each model the residual sums of squares $D = \sum \epsilon^2$, 
where the sum is taken over fitted data points.  Under the null hypothesis that the extra complexity is not significant, and assuming the data points to be independent, the test statistic defined by 
\begin{equation}
F = \frac{(D_1 - D_2)/(d_1 - d_2)}{D_2/d_2}, \label{Fstatistic}
\end{equation}
(where $d_1$ and $d_2$ count the residual degrees of freedom of $m_1$ and $m_2$ respectively) has an F distribution with degrees of freedom equal to $d_1 - d_2$ and $d_2$.  This allows us to assign a p-value to the significance of model $m_2$ over $m_1$.  

\begin{table}
\caption{\small P-values testing significance of the extra parameter related to nested pairs of models in Eqs.~\ref{alpha_m0} to \ref{alpha_m5}.  Smaller p-values indicate that the extra parameters in the more complicated model are significant.  The second column gives the extra parameters included in the more complicated of the two models.  
}
\label{tab:alpha_Pvalues}
\begin{center}
\begin{tabular}{lcll}
\hline
      &  Parameter &  Dataset I & Dataset II \\
\hline
model 0 to model 1:  & $c_2$                            &  $< 2\times 10^{-16}$  &  $< 2\times 10^{-16}$  \\
model 1 to model 2:  & $c_3$                            &  $1.2\times 10^{-  9}$  &  $1.9\times 10^{-  6}$  \\
model 2 to model 3:  & $c_4$                            &  0.0098                           & 0.648  \\
model 1 to model 4:  & $\lambda_\alpha, \mu_\alpha$ & $1.1\times 10^{-10}$  &  0.60\\
model 2 to model 5:  & $\lambda_\alpha, \mu_\alpha$ & $3.4\times 10^{-  5}$  &  0.81\\
model 4 to model 5:  & $c_3$                            & 0.00064                          & $2.7\times 10^{-6}$ \\
\hline
\end{tabular}
\end{center}
\end{table}

We have fitted each of the five models to the combination $\alpha = (A - a)/b$ using $a$ and $b$ from Table~\ref{tab:parameters} and $A$ from the hyperbolic response function fits of both PM and MM data for datasets I and II separately.  The number of data points fitted, that is, the number of fitted hyperbolic response functions for which all three parameters $A$, $B$ and $K$ are positive, was 364 for dataset I and 460 for dataset II.  Table~\ref{tab:alpha_Pvalues} gives the calculated p-values.  

The parameters $c_2$ and $c_3$ modelling linear effects in $\Delta G^{\rm DNA/RNA}$ and $n_{\rm pyr}$ respectively are highly significant in both datasets.  The parameter $c_4$ defining a mixed effect is barely significant at the 1\% level in dataset I and not significant in dataset II, and we shall ignore it from here on.  

The probe folding effect is highly significant for dataset I, but not significant for dataset II.  Note that Models~4 and 5 contain the functional form 
\begin{equation}
-\ln\{1 + \exp[\lambda(\mu - \Delta g)]\} \approx \left\{ 
			\begin{array}{lr}  \lambda(\Delta g - \mu), & \Delta g << \mu, \\
			                              0,                                        &\Delta g >> \mu, \end{array} \right.  \label{deltaGlimits}
\end{equation}
for $\lambda > 0$.  Thus the probe folding effect ``switches on'' once the energy of a folded probe is below some threshold $\mu_\alpha$, at which point the effect becomes linear in $\Delta g^{\rm DNA-fold}$.    From Table~\ref{tab:parameters}, the fitted value of $\mu_\alpha$ in Model~5 for dataset I is 4.57, whereas the range of probe folding energies calculated by Mfold for the probe sequences of the U95 microarray is $-8.18 < \Delta g^{\rm DNA-fold} < 2.98$.  Thus $\mu_\alpha$ is well above the folding energy of any of the probes, implying that the probe folding effect is effectively linear for dataset I.  On the other hand, the fitted value of $\mu_\alpha$ in Model~5 for dataset II is $-6.14$, which is below the range $-5.49 < \Delta g^{\rm DNA-fold} < 3.11$ of calculated  probe folding energies for the U133 microarray, implying that the probe folding is switched off for dataset II.  There is no obvious reason why the probe folding parameter $\mu_\alpha$ should should shift markedly from one spike-in experiment to another, given that, although the U133 probeset nucleotide sequences are completely redesigned from those of the U95 microarray, the experimental protocol and geometry of the microarray should be similar.  One, albeit prosaic, explanation may simply be that dataset II is noisier (see Table~\ref{tab:coef_of_var}) and the probe folding effect has been lost in the noise.  

\begin{figure}
\includegraphics[scale=0.55]{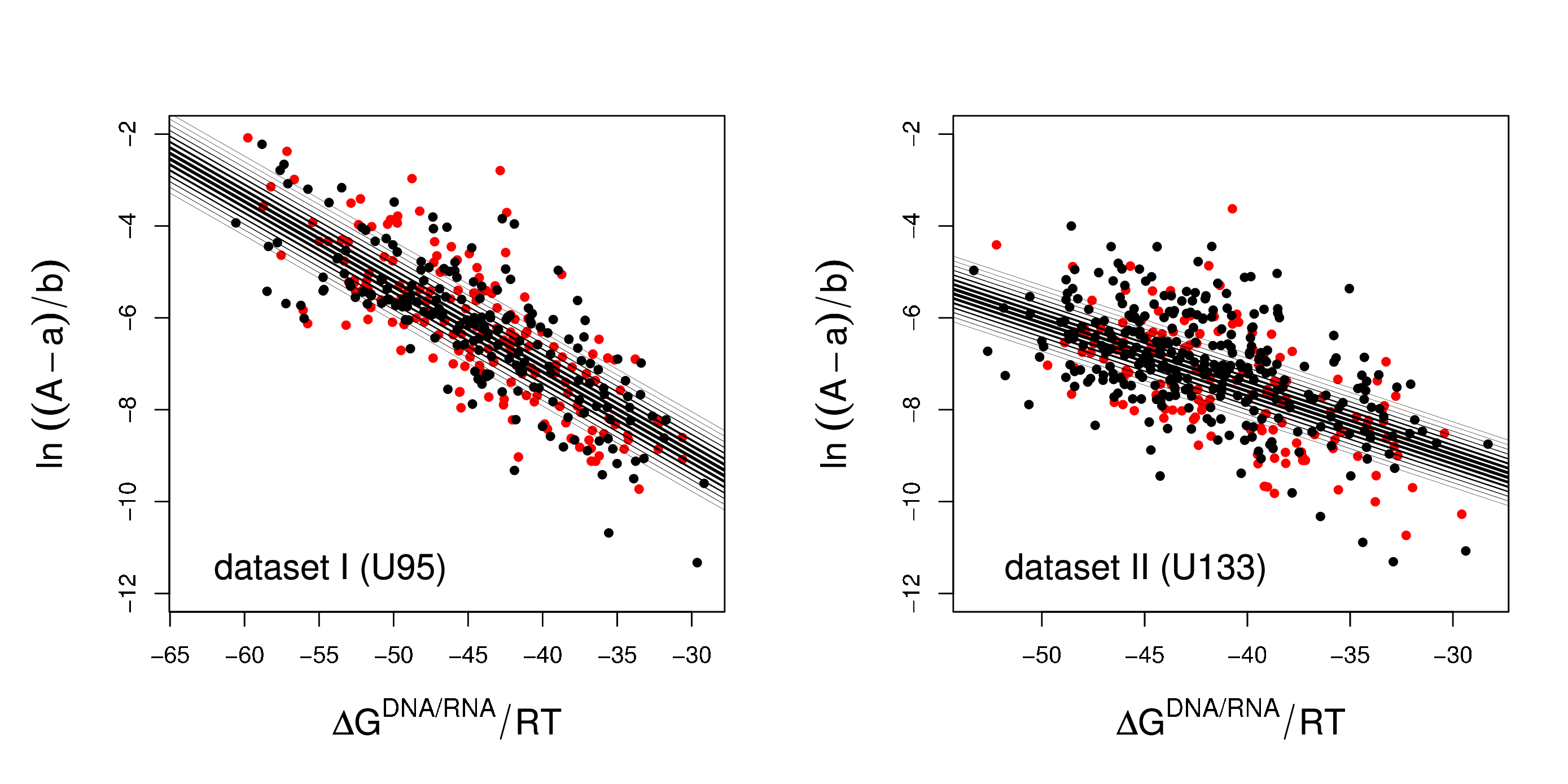}
\centering
\caption{Fits of Eq.~\ref{alpha_m2} (Model 2) to the parameter $A$ of the hyperbolic response function fits of both PM (black) and MM (red) data for datasets I and II.  The parameters $a$ and $b$ are from Table~\ref{tab:parameters}.  The plotted lines correspond to the range $6 \le n_{\rm pyr} \le 20$ of pyrimidine counts, with $n_{\rm pyr}$ increasing from bottom to top.}
\label{fig:A_fit}
\end{figure}

Fitted parameter values of Models~2 and 5 are given in Table~ \ref{tab:parameters}.  As expected from the above argument, the fitted parameters $c_1$, $c_2$ and $c_3$ for dataset II differ very little between Models~2 and 5.  Fits of Model~2 to the data are shown in Fig.~\ref{fig:A_fit}.

\subsection{Horizontal scale parameter}
\label{sec:Horizontal}

The parameter $K$ sets the horizontal scale of the hyperbolic response function Eq.~\ref{hyperbolicResponse}.  $K^{-1}$ is an estimate of the spike-in concentration required to give a fluorescence intensity half way between the background, zero concentration level and the asymptotic, infinite concentration level.  Our model in Section~\ref{sec:Model} explains $K$ as an effective rate reaction constant which is determined by the reactions occurring during the hybridisation step, and which is unaffected by the washing step.  As was the case for the vertical scale parameters, much of the information required to evaluate $K$ from first principles is unknown, and so we try for an ansatz based on probe sequences and free binding energies.  

Eqs.~\ref{KModel}, \ref{XbulkDef} and \ref{XsurfaceDef} give $K$ in terms of reaction rate constants and concentrations of reactants.  In general, each term $K^r$ or $X^r$ occurring in Eq.~\ref{KModel}, where $r$ labels one of the reactions in Table~\ref{tab:reactions}, is a sum of terms of the form $\mbox {const.} \times e^{-\Delta G^r_i /RT}$ , weighted by the concentration of reactant $i$.  Once again we will be guided by Heim et al.'s numerical estimate of $X^\bulk$~\cite{Heim06}, and approximate each sum as a single exponential term.  Thus we write 
\begin{equation}
K^r \mbox{ or } X^r =  \exp[\lambda_r(\mu_r - \Delta g^r)], 
\end{equation}
where the $\mu_r$ and $\lambda_r$ are fitting parameters, and $\Delta g^r = \Delta G^r/RT$, with the effective binding energy $\Delta G^r$ for each reaction is estimated from some external physical model.   With the sign convention that $\Delta G^r$ is defined negative for a bound state, each $\lambda_r$ is expected to be positive.  

Consider first dataset III, for which the complex non-specific background is absent.  In Eq.~\ref{KModel} we can set the non-specific binding terms $X^\bulk$ and $X^\NS$ to zero, giving 
\begin{equation}
\ln K = \ln K^\spec - \ln(1 + K^\Sfold) - \ln(1 + K^\Pfold). 
\end{equation} 
The rate constants are modelled as 
\begin{eqnarray}
K^\spec & = & \exp[\lambda_\spec(\mu_\spec - \Delta g^{\rm DNA/RNA})], \label{KSansatz}\\
K^\Sfold & = & \exp[\lambda_\Sfold(\mu_\Sfold - \Delta g^{\rm RNA/RNA})], \\
K^\Pfold & = & \exp[\lambda_\Pfold(\mu_\Pfold - \Delta g^{\rm DNA-fold})]. 
\end{eqnarray}
For the free binding energy $\Delta G^{\rm DNA/RNA}$ we use the nearest neighbour stacking model parameters of Sugimoto et al.\cite{Sugimoto95}, for $\Delta G^{\rm RNA/RNA}$ we use Xia et al.'s nearest neighbour stacking parameters for RNA binding to RNA~\cite{Xia98}, and for $\Delta G^{\rm DNA-fold}$ we use Zuker's Mfold web server~\cite{Zuker03}.  The Mfold web server also has the facility to calculate folding energies of RNA targets.  However, since for RNA target folding we are interested in the propensity for the 25-mer stretch of target complimentary to a given probe to bind with any segment of the much longer target RNA (or possibly another RNA molecule), we believe it is more appropriate to model target folding in solution using an RNA-to-RNA binding energy.  

\begin{figure}
\includegraphics[height=8cm]{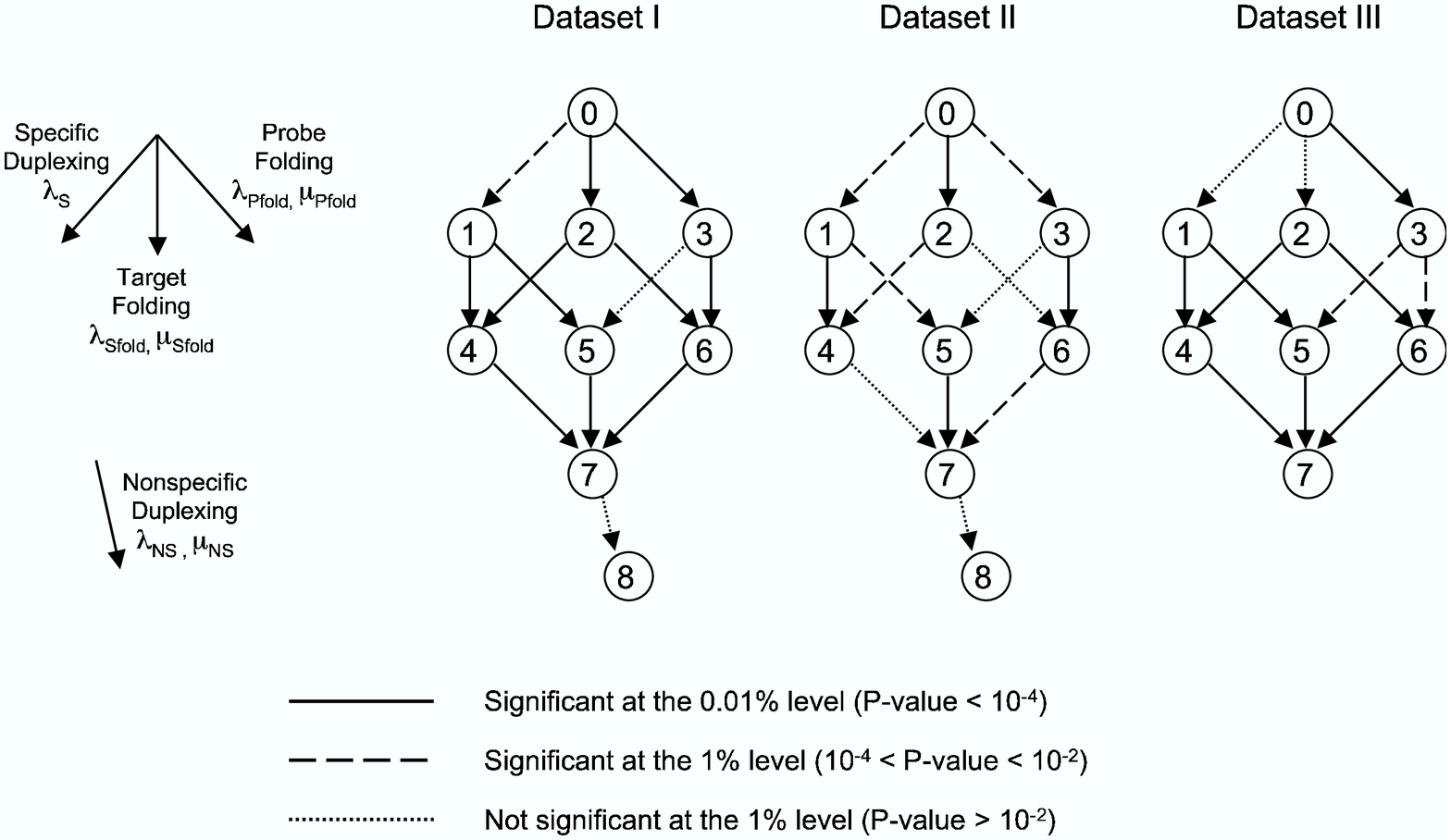}
\centering
\caption{The relationship between the nine models fitting the effective rate constant $K$, namely Eqs.~\ref{model_K0} to \ref{model_K7} and \ref{model_K8}.  The directions of the arrows indicate the extra effects included in going from a simpler, nested model to a more complicated model, and the style of the arrows indicate P-values from Table~\ref{tab:K_Pvalues}.  Smaller P-values indicate that the extra parameters in the more complicated model are significant.  The fitted parameters corresponding to the favoured model, Model 7, are given in Table~\ref{tab:parameters}}
\label{fig:K_models}
\end{figure}

To try to understand the relative importance of each of the effects contributing to the effective rate constant $K$ we have analysed a set of models containing nested pairs, the relationship between which is illustrated in Fig.~\ref{fig:K_models}: 
\begin{eqnarray}
\mbox{Model~0:  }  \ln K & =  & k_0 + \epsilon, \label{K_m0}  \label{model_K0} \\
\mbox{Model~1:  } \ln K & =  &  \lambda_\spec(\mu_\spec - \Delta g^{\rm DNA/RNA}) + \epsilon,  \label{model_K1} \\
\mbox{Model~2:  }  \ln K & =  &  k_0 - \ln\{1 + \exp[\lambda_\Sfold(\mu_\Sfold - \Delta g^{\rm RNA/RNA})]\}  + \epsilon,   \label{model_K2} \\
\mbox{Model~3:  } \ln K & =  &  k_0 - \ln\{1 + \exp[\lambda_\Pfold(\mu_\Pfold - \Delta g^{\rm DNA-fold})]\}  + \epsilon,   \label{model_K3} \\ 
\mbox{Model~4:  }  \ln K & =  &  \lambda_\spec(\mu_\spec - \Delta g^{\rm DNA/RNA}) \nonumber \\
                      & &  - \ln\{1 + \exp[\lambda_\Sfold(\mu_\Sfold - \Delta g^{\rm RNA/RNA})]\}  + \epsilon,   \label{model_K4} \\ 
\mbox{Model~5:  } \ln K & =  &  \lambda_\spec(\mu_\spec - \Delta g^{\rm DNA/RNA}) \nonumber \\
                      & &  - \ln\{1 + \exp[\lambda_\Pfold(\mu_\Pfold - \Delta g^{\rm DNA-fold})]\}  + \epsilon,   \label{model_K5} \\
\mbox{Model~6:  }  \ln K & =  &  k_0 - \ln\{1 + \exp[\lambda_\Sfold(\mu_\Sfold - \Delta g^{\rm RNA/RNA})]\}  \nonumber \\
                      & &  - \ln\{1 + \exp[\lambda_\Pfold(\mu_\Pfold - \Delta g^{\rm DNA-fold})]\}  + \epsilon,   \label{model_K6} \\
\mbox{Model~7:  } \ln K & =  &  \lambda_\spec(\mu_\spec - \Delta g^{\rm DNA/RNA}) \nonumber \\
                      & &  - \ln\{1 + \exp[\lambda_\Sfold(\mu_\Sfold - \Delta g^{\rm RNA/RNA})]\} \nonumber \\ 
                      & &  - \ln\{1 + \exp[\lambda_\Pfold(\mu_\Pfold - \Delta g^{\rm DNA-fold})]\}  + \epsilon, \label{model_K7}
\end{eqnarray}
where $\epsilon$ is the residual error, which is assumed Gaussian.  The first term in each of Models~1, 4, 5 and 7 could equally well be written as $k_0 + k_1 \Delta g^{\rm DNA/RNA}$ to make the nesting explicit, but for convenience we use a parameterisation based on Eq.~\ref{KSansatz}.  Models~1 to 3 include only the effects of specific hybridisation, target folding in bulk solution and probe folding respectively.  Models~4 to 6 include pairwise effects, and Model~7 includes all three effects.  A recurring theme in these models is the functional form of Eq.~\ref{deltaGlimits}.  Thus the target and probe folding effects ``switch on'' once the binding energy is below (i.e. more tightly binding than) thresholds $\mu_\Sfold$ and $\mu_\Pfold$ respectively, and have the effect of reducing the effective rate constant $K$.  

\begin{table}
\caption{\small P-values testing significance of the extra parameters related to nested pairs of Models 0 to 8 fitting the effective rate constant $K$.  Smaller p-values indicate that the extra parameters in the more complicated model are significant.  The number of fitted PM hyperbolic response functions for which all three parameters $A$, $B$ and $K$ positive, and hence the number of points fitted to the models, is 188 for dataset I, 303 for dataset II and 192 for dataset III.  
}
\label{tab:K_Pvalues}
\begin{center}
\begin{tabular}{lclll}
\hline
      & Parameter &  Dataset I & Dataset II & Dataset III \\
\hline
model 0 to model 1:  &  $\lambda_\spec$                      &   $0.00028$                   		&     $0.00021$                   &     $0.51$    \\
model 0 to model 2:  &  $\lambda_\Sfold, \mu_\Sfold$   &  $3.5\times 10^{-11}$    	&     $1.1\times 10^{-7}$    &     $0.32$    \\
model 0 to model 3:  &  $\lambda_\Pfold, \mu_\Pfold$   &   $1.6\times 10^{-12}$    	&     $0.00011$                   &     $4.8\times 10^{-15}$    \\
model 1 to model 4:  &  $\lambda_\Sfold, \mu_\Sfold$   &   $< 2\times 10^{-16}$    	&     $2.3\times 10^{-6}$    &     $1.8\times 10^{-9}$    \\
model 1 to model 5:  &  $\lambda_\Pfold, \mu_\Pfold$   &   $3.0\times 10^{-10}$    	&     $0.0077$                      &     $< 2\times 10^{-16}$    \\
model 2 to model 4:  &  $\lambda_\spec$                       &   $4.4\times 10^{-12}$    	&     $0.0055$                      &     $5.9\times 10^{-10}$    \\
model 2 to model 6:  &  $\lambda_\Pfold, \mu_\Pfold$   &   $6.8\times 10^{-7}$    	&     $0.089$                        &     $< 2\times 10^{-16}$    \\
model 3 to model 5:  &  $\lambda_\spec$                       &   $0.087$                  		&     $0.023$                        &     $0.00016$    \\
model 3 to model 6:  &  $\lambda_\Sfold, \mu_\Sfold$   &     $1.4\times 10^{-5}$    	&     $9.6\times 10^{-5}$    &     $0.00012$    \\
model 4 to model 7:  &  $\lambda_\Pfold, \mu_\Pfold$   &  $9.1\times 10^{-5}$    	&     $0.056$                        &     $7.7\times 10^{-14}$    \\
model 5 to model 7:  &  $\lambda_\Sfold, \mu_\Sfold$   &    $3.8\times 10^{-13}$    	&     $1.7\times 10^{-5}$    &     $2.0\times 10^{-5}$    \\
model 6 to model 7:  &  $\lambda_\spec$                       &  $7.9\times 10^{-10}$    	&     $0.0034$                      &     $2.4\times 10^{-5}$    \\
model 7 to model 8:  &  $\lambda_\NS, \mu_\NS$   	   &     $0.016$   			&     $0.23$                           &       \\
\hline
\end{tabular}
\end{center}
\end{table}

\begin{figure}
\includegraphics[scale= 0.53]{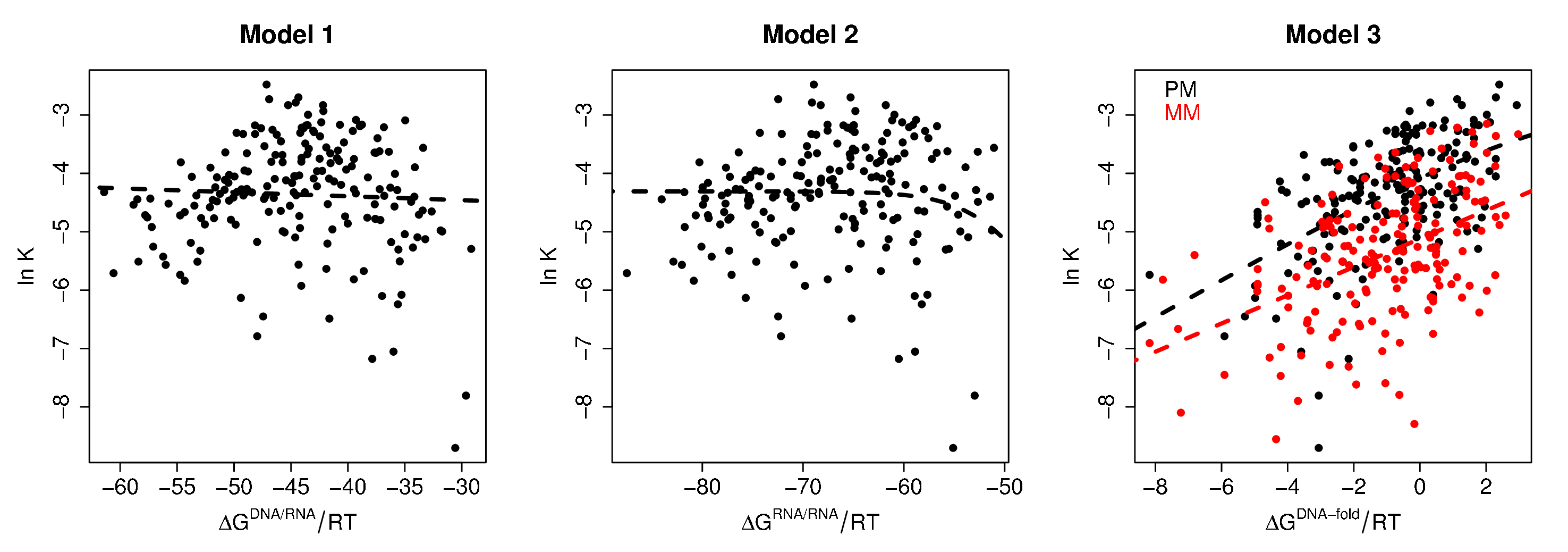}
\centering
\caption{Fits of $\ln K$ estimated from Dataset III to Models 1, 2 and 3.  Mismatch data is also shown for Model 3 since $\Delta G^{\rm Pfold}$ can be obtained from the Mfold web site for all probe sequences.}
\label{fig:K_nbg_models}
\end{figure}

P-values calculated from the F-statistic, Eq.~\ref{Fstatistic}, testing the pairwise comparative significance of these models are given for Dataset III in the right hand column of Table~\ref{tab:K_Pvalues} (see also Fig.~\ref{fig:K_models}).  Results are shown for PM data only as no complete set of stacking model parameters for $\Delta G^{\rm DNA/RNA}$ with mismatches is available.  Comparisons of Model 0 with Models 1 to 3 indicate that, taken in isolation,  the specific hybridisation and bulk target folding effects appear not to be significant, whereas the probe-folding effect appears to be highly significant.  This is also illustrated in Fig.~\ref{fig:K_nbg_models}.  However, when taken in combination with probe folding, the analysis shows specific binding and target folding to be significant at the 1\% level (see the comparisons Model 3 to 5 and 3 to 6 in Table~\ref{tab:K_Pvalues}).  Thus we accept Model 7 for Dataset III.  The fitted parameters are given in the right hand column of Table~\ref{tab:parameters}.  Note that each $\lambda_r$ is positive as required of the model.  

For Datasett III, the apparent non-significance of the specific hybridisation and bulk target folding effects in Models~1and 2 can be explained as follows.  In Model~7, the bulk folding is a stronger effect than specific hybridisation by a factor of 2 ($\lambda_\Sfold \approx 2\lambda_\spec$ in Table~\ref{tab:parameters}).  Furthermore, from Eq.~\ref{deltaGlimits}, the bulk folding effect is opposite in sign to the specific hybridisation effect, and only comes into effect for $\Delta g^{\rm RNA/RNA} < \mu_\Sfold = -73.7$. Also, it turns out that $\Delta g^{\rm DNA/RNA}$ and $\Delta g^{\rm RNA/RNA}$ are very highly correlated, with a Pearson correlation coefficient of 0.89.  Examination of the first two plots in Fig.~\ref{fig:K_nbg_models} shows a tendency of the data to increase with $\Delta g$ to start with, while the bulk target folding dominates, and then to decrease once the bulk folding effect switches off and the specific hybridisation effect takes over.  Attempting to fit a straight line through data which first increases and then decreases has resulted in the conclusion that the term linear in $\Delta G$ in Model~I is not significant.  A related statement has been made by Carlon and Heim~\cite{Carlon06}, namely that the effective target concentration needs to be appropriately ``rescaled'' for those targets with a high binding affinity in bulk solution in order to see the expected relationship between $K$ and $\Delta G^{\rm DNA/RNA}$.  

\begin{figure}
\includegraphics[scale= 0.53]{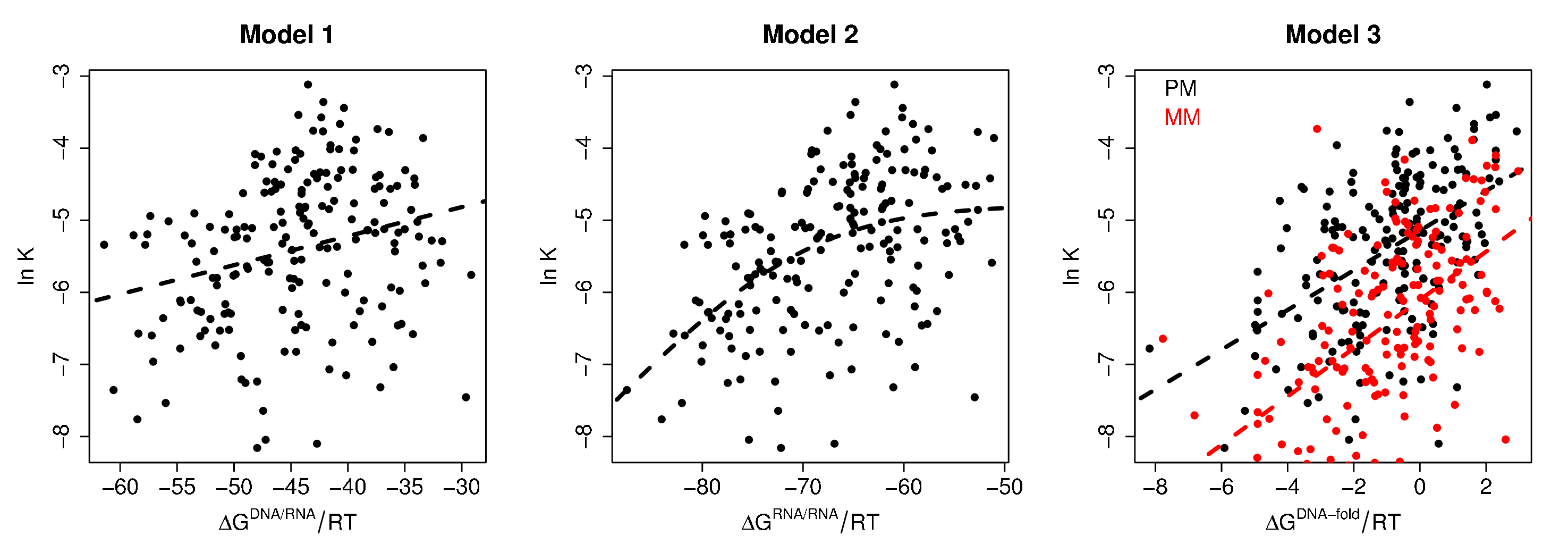}
\centering
\caption{The same as Fig.~\ref{fig:K_nbg_models} for Dataset I.}
\label{fig:K_datasetI_models}
\end{figure}

We now turn to Datasets I and II.  In the presence of a complex non-specific background, $X^\bulk$ and $X^\NS$ are reinstated in Eq.~\ref{KModel}.  The bulk hybridisation effect will be a sum of exponentials of $\Delta G_i^{\rm RNA/RNA}$, and its modelling can be absorbed into that for bulk target folding, while the non-specific effect will be a sum of  exponentials of $\Delta G_i^{\rm DNA/RNA}$.  Thus we set 
\begin{eqnarray}
K^\Sfold + X^\bulk & = & \exp[\lambda_\Sfold(\mu_\Sfold - \Delta g^{\rm RNA/RNA})], \\
X^\NS & = & \exp[\lambda_\NS(\mu_\NS - \Delta g^{\rm DNA/RNA})], 
\end{eqnarray}
which suggests one further model: 
\begin{eqnarray}
\mbox{Model~8:  } \ln K & =  & \lambda_\spec(\mu_\spec - \Delta g^{\rm DNA/RNA}) \nonumber \\
                      & &  - \ln\left\{1 + \exp[\lambda_\Sfold(\mu_\Sfold - \Delta g^{\rm RNA/RNA})]\right\} \nonumber \\ 
                      & &  - \ln\left\{1 + \exp[\lambda_\Pfold(\mu_\Pfold - \Delta g^{\rm DNA-fold})]\right. \nonumber \\
                      & & \qquad \left. + \exp[\lambda_\NS(\mu_\NS - \Delta g^{\rm DNA/RNA})]\right\}  + \epsilon.   \label{model_K8}
\end{eqnarray}
Turning to Table~\ref{tab:K_Pvalues}, columns I and II, we discover that the extra parameters introduced to account for non-specific probe-target binding are not significant at the 1\% level (see also Fig.~\ref{fig:K_models}).  This surprising result can be explained by the fact that the fitted values of $\mu_\NS$ are in both cases close to the maximum value of $\Delta g^{\rm DNA/RNA}$ within the dataset, so most of the fitted points fall into the  $\Delta g^{\rm DNA/RNA} << \mu_{\rm NS}$ regime of Eq.~\ref{deltaGlimits}, and the effect is adequately covered by the $\lambda_\spec$ term of Model 7.  To further illustrate the point, fits to Models 1, 2 and 3 are plotted In Fig.~\ref{fig:K_datasetI_models}.  If Model 1 is taken in isolation, $\lambda_\spec$ appears to have the ``wrong'' sign, as the non-specific probe-target binding and target folding and binding in solution all combine to dominate the specific binding effect.  A similar result is observed for dataset II.  

The generally small p-values in the first column of Table~\ref{tab:K_Pvalues} indicate that Model 7 is an appropriate description of the parameter $K$ for dataset I. For dataset II the picture is less clear.  In agreement with the analysis of the parameter $\alpha$, the probe folding is in general less significant.  Nevertheless, for consistency we list the fitting parameters of Model 7 to both datasets in Table~\ref{tab:parameters}, while acknowledging there is redundancy in the Dataset II parameters.  

\subsection{How close are the fits?} 

\begin{figure}
\includegraphics[scale=0.52]{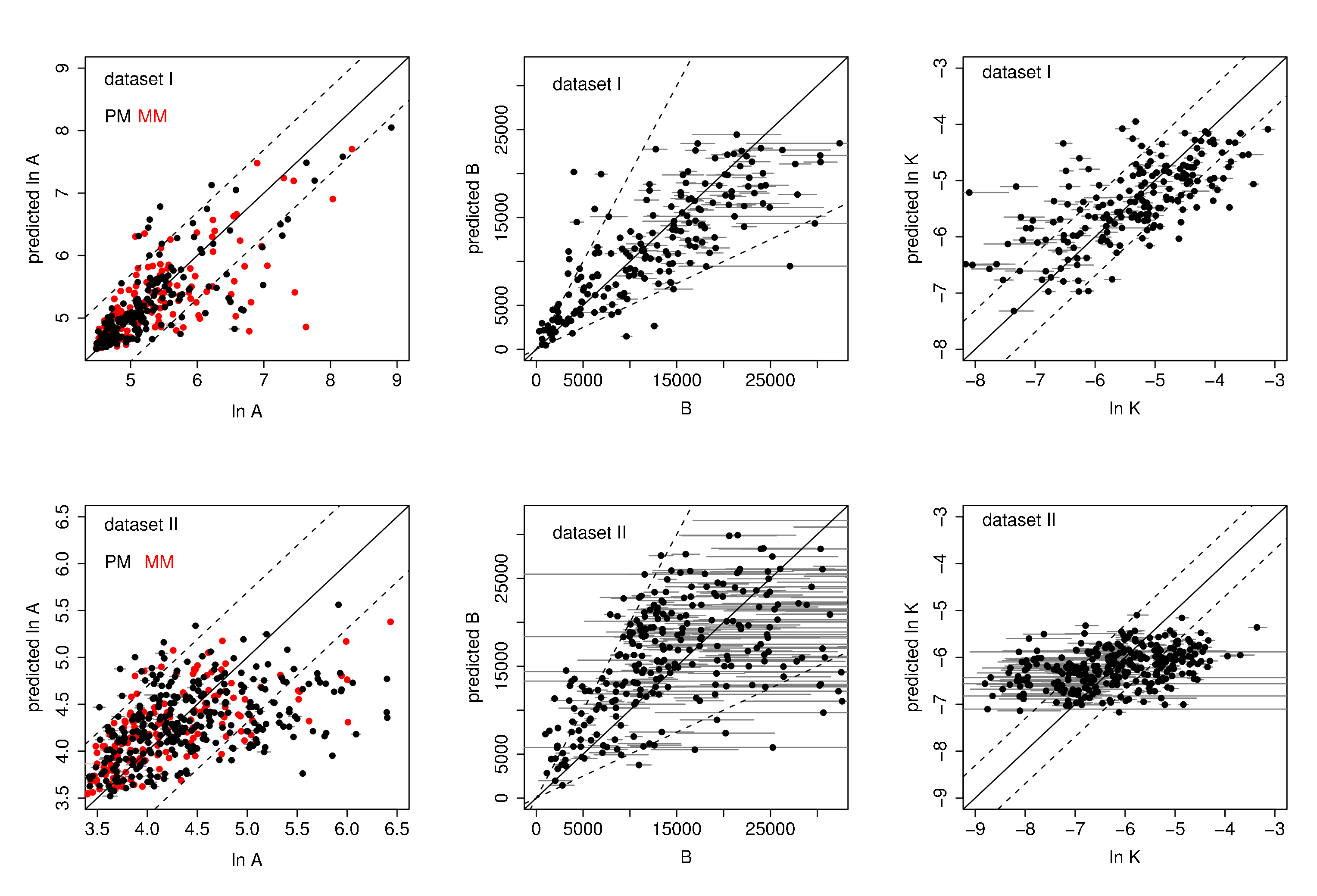}
\centering
\caption{Estimated parameters obtained by fitting the the hyperbolic response function Eq.~\ref{hyperbolicResponse} to datasets I and II (horizontal axes, together with error bars showing standard errors) plotted against with the values that would be predicted by the quantitative fits of Section~\ref{sec:Quantitative} (vertical axes).  The dotted lines indicate a factor of 2 either side of the diagonal.  }
\label{fig:test_fits}
\end{figure}

Fig.~\ref{fig:test_fits} gives some idea of how much information has been lost in the above fits.  The plot compares estimated parameters $A$, $B$ and $K$ of the fitted hyperbolic response curves, such as those in Fig.~\ref{hyperbolicResponse}, with those which would be predicted by the fitting constants and parameters listed in Table~\ref{tab:parameters}, namely 
\begin{eqnarray}
A & = & a + be^{c_1 + c_2\Delta g^{\rm DNA/RNA} + c_3n_{\rm pyr}- \ln\{1 + \exp[\lambda_\alpha(\mu_\alpha - \Delta g^{\rm DNA-fold})]\}}, 
						\label{AfromTab4}    \\ 
B & = & a + be^{-c_0 \exp\left(\lambda_0 \Delta g^{\rm RNA/RNA}\right)} - A, 
						\label{BfromTab4}    \\ 
K & = & \frac{ e^ {\lambda_\spec(\mu_\spec - \Delta g^{\rm DNA/RNA})}}
                     { \left[1 + e^{\lambda_\Sfold(\mu_\Sfold - \Delta g^{\rm RNA/RNA})} \right] 
                      \left[1 + e^{\lambda_\Pfold(\mu_\Pfold - \Delta g^{\rm DNA-fold})} \right] }.  \label{KfromTab4} 
\end{eqnarray}
Dotted lines either side of the diagonal are the boundary of the region within which predicted values 
do not differ from the original fitted parameters by more than a factor of 2.  A clear majority of estimates of $A$ and $B$ fall within this range.  Clearly the most difficult parameter to explain adequately is the horizontal scale $K$, owing to the large number of contributing chemical reactions.  In general, dataset II has proved to be more problematic than dataset I, probably because the concentration range tested does not extend far enough into the saturation regime to demonstrate a clear hyperbolic isotherm.   

\section{Parameter prediction}
\label{sec:Prediction}

For the above model to be of value in constructing a practical algorithm for inferring target concentrations, some or all of its parameters should ideally be predictable using only information available to experimental biologists.  That available information consists of fluorescence intensities for the complete set of features on each microarray used in an experiment, the probe sequences of each feature, and any parameters associated with the experimental protocol.  By contrast, the fitted parameters of Table~\ref{tab:parameters} were obtained from spike-in experiments.  Comparing datasets I and III in Figs.~\ref{fig:A_comparison} and \ref{fig:K_comparison}, one sees for instance that the unknown nature of the complex background has a profound effect on the parameters $A$ and $K$ of the hyperbolic response function.  At first sight it appears one may need a new set of spike-in data for each experiment, which is clearly not a practical consideration.  However, we argue here that if one exploits the distribution of fluorescence intensities from the entire microarray, an estimation of vertical scale parameters at least may be possible.  

In the following qualitative description we propose a two step process for the vertical scale parameters, in which the physical background $a$ and maximum intensity $b$ for a microarray are first determined from the entire distribution of intensities over the microarray.  The intensities $I(x)$ can then be scaled to the dimensionless coverage fraction $\theta(x)$ via Eq.~\ref{intensity}, and one is left with the remaining problem of estimating the parameters $\alpha$ and $\beta$, which are driven by the chemical reactions of Table~\ref{tab:reactions}.  

To estimate $a$ and $b$, consider the histograms in Fig.~\ref{fig:histogram}.  For both datasets I and II, our estimate of the physical background $a$, based on hyperbolic response curves derived from spike-in data, is close to two standard deviations above the minimum measured fluorescence intensity.  Assuming an experiment consisting of a number of technical replicates of each hybridisation setup, the data can be quantile normalised across replicates.  A representative minimum intensity $a_{\rm min}$ can be obtained by fitting a suitable smooth curve to the logged histogram (i.e. the lower panel of Fig.~\ref{fig:histogram}), and the coefficient of variation in the data $\eta$ easily estimated from the replicate intensity values over the whole microarray.  $(1 + 2\eta)a_{\rm min}$ then gives an estimate of $a$.  

Estimating $b$ from the histogram proves to be quite difficult because of the gradual tail at its the right hand end.  With some experimentation we find that a cubic fit to the logged histogram over the range $[l + 0.25(u - l), l + 0.875(u - l)]$, where $l$ and $u$ are the lower and upper extremities of the histogram, crosses the $\log_{10}(\mbox{count}) = 0$ line close to two standard deviations above the previously obtained estmate of $a + b$.  Calling this point $(a + b)_{\rm max}$, an estimate of $a + b$ is then $(1 - 2\eta)(a + b)_{\rm max}$.  However, we find that such a method is highly sensitive to the range over which the cubic is fitted.  

\begin{figure}
\includegraphics[scale=0.46]{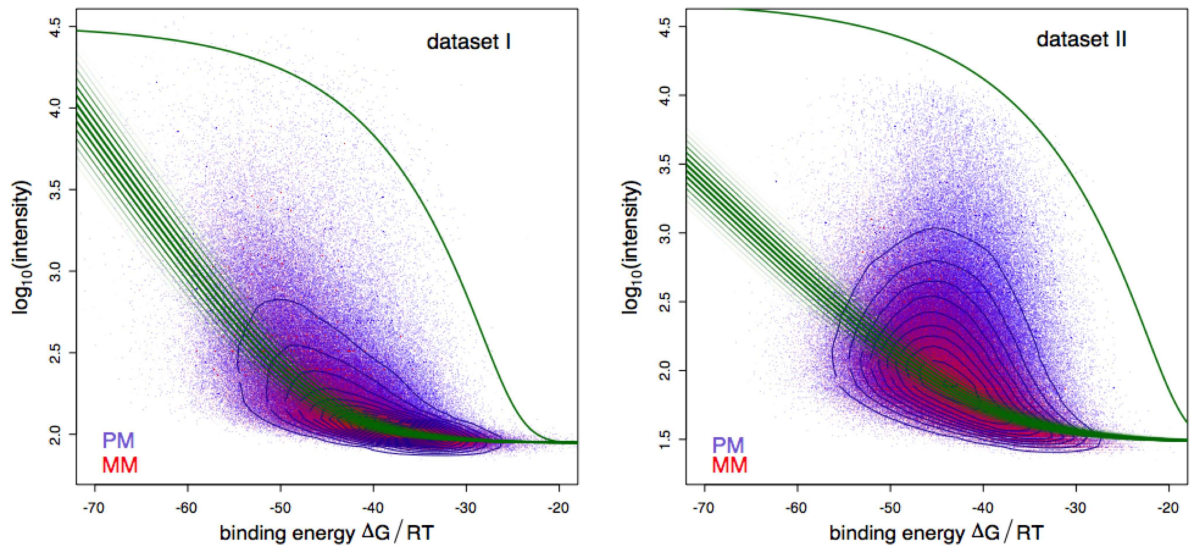}
\centering
\caption{Scatter plots of measured fluorescence intensities against the theoretical DNA/RNA free binding energies.  The upper curve is the fit to $A + B$ of Eq.~\ref{BfromTab4}, and the lower set of curves are the fits to $A$ of Eq.~\ref{AfromTab4} for pyrimidine counts $6 < n_{\rm pyr} < 20$, with $n_{\rm pyr}$ increasing from bottom to top.  Also shown are contour lines of the density of points.  }
\label{fig:intensity_scatter_plot}
\end{figure}

To gain some insight into how $\alpha$ and $\beta$ may be estimated, consider the scatter plot, Fig.~\ref{fig:intensity_scatter_plot}, of fluorescence intensities against the theoretical free binding energy $\Delta G^{\rm DNA/RNA}$ obtained from the probe letter sequences using the nearest neighbour stacking model~\cite{Sugimoto95}.  Superimposed on these plots are the fits from Eqs.~\ref{AfromTab4} and \ref{BfromTab4} to the asymptotic saturation intensity $A + B$ and background intensity at zero spike-in concentration $A$, using the parameters of Table~\ref{tab:parameters}.  According to our model, the asymptote curve should form an upper envelope to the data, with some slight leakage across the envelope due to the finite coefficient of variation in the data.  Because the vast majority of the genes are not expressed in RNA samples taken from a typical cell, most of the data is expected to lie along, or close to, the lower set of background intensity curves.  Indeed this is precisely what is seen.  Conversely, in the absence of spike-in data, there is potential to estimate the upper, asymptote intensity curve by fitting an envelope to the data and the lower, background intensity curve by fitting a curve through the ridge of the scatter plot's contour lines.  In principle, these fitted curves, together with $a$ and $b$ then determine estimates of $\alpha$ and $\beta$ for each feature on the microarray.  

\section{Conclusions and outlook}
\label{sec:Conclusions}

This paper concentrates mainly on the immediate aim of understanding the physical processes at work in the operation of microarrays, but in doing so highlights some of the challenges and, we hope, gives some guidance, for meeting the ultimate aim of providing an algorithm for converting the set raw fluorescence intensities from microarrays to absolute target concentrations.   

The model we have examined includes the effects of specific and non-specific hybridisation, folding and hybridisation in bulk solution of target RNA, the folding of probes at the microarray surface, and the removal of signal during the post-hybridisation step.  It leads to the hyperbolic response curve (or Langmuir isotherm) of Eq.~\ref{hyperbolicResponse} with three fitting parameters $A$, $B$ and $K$, which depend on a set underlying physical physical parameters including chemical reaction rate constants, washing survival functions and RNA target concentrations.  In more practical terms, all three parameters will depend on the probe letter sequence, whether the probe is PM or MM, the nature of the complex non-specific background, and experimental protocols such as hybridisation temperature and washing times.  Determining the parameters only from information likely to be known to biologists in a practical situation, as opposed to a highly controlled spike-in experiment, remains a formidable task.  

The model is tested against the Affymetrix U95 spike-in datasets with and without a complex non-specific background and the Affymetrix U133 spike-in dataset.  In general, agreement with a hyperbolic response curve is excellent for the U95 spike-in datasets and reasonable for the U133 spike-in dataset (see Table~\ref{tab:coef_of_var}).   

The response function parameters $A$ and $B$ set the vertical scale of the isotherm, that is, the scale of the measured fluorescence intensities.  The parameter $A$ is a combination of a relatively straightforward physical background, and a non-trivial contribution from the complex non-specific background.  It is important to understand the nature of the non-specific background component as it is responsible for the ``bright mismatches'' problem which complicates the naive PM $-$ MM subtraction scheme used in the MAS5 algorithm, for instance, for dealing with non-specific hybridisation.  Our analysis shows that the DNA/RNA binding energy, pyrimidine content, and (in the case of U95 dataset) the folding of probes, all contribute significantly to the value of this parameter.  The dependence of $A$ on binding energies and pyrimidine count is illustrated in Fig.~\ref{fig:A_fit}.

The parameter $B$ (or, more precisely, the combination $A + B$, where $B >> A$ in general) is mainly concerned with the asymptote at high spike-in concentration, which, according to our model, is driven by the washing step.  The qualitative prediction that it should be less for a mismatch feature than for a perfect match feature in a PM/MM pair is verified for all three datasets in Fig.~\ref{fig:AplusB_comparison}.  Its quantitative behaviour as a function of specific probe-target binding energies, using bulk solution free binding energies as a guide, is verified in Fig.~\ref{fig:AplusB_fit}.  

The parameter $K$ can be thought of a an effective overall reaction rate.  It sets the horizontal scale of the isotherm, that is, the scale of the specific target concentration.  If an algorithm to determine absolute target concentrations is to be constructed, it is necessary to understand this parameter.  Because of the large number of hybridisation reactions which have the potential to contribute it is by far the hardest of the three parameters to explain effectively.  The model predicts that it is affected by all of the reactions listed in Table~\ref{tab:reactions} occurring in the bulk hybridisation solution and at the microarray surface, but is unaffected by the dissociation during the washing phase.  Analysis to determine which reactions are significant is complicated by the fact that the effect on $K$ of non-specific hybridisation and probe and target folding act in the opposite direction to that of specific binding, so obscuring the effects.  Nevertheless, our analysis indicates that all of the hybridisation reactions considered are potentially significant contributors.  

A common practice in previous studies~\cite{Hekstra03,Held03,Burden04,Carlon06,Heim06,Held06} has been to invert fits of hyperbolic response curves to recover spike-in target concentrations in order to test the predictive ability of models.  We have deliberately refrained from doing so here, as one of the results of this study has been to demonstrate the strong dependence of the parameters of the isotherm on the (in practice unknown) complex non-specific background.  Recovering spike-in concentrations using fitting parameters which implicitly contain information about the background belonging to a particular dataset is an inherently circular argument and is guaranteed to give unrealistically good results.   

Instead, in Section~\ref{sec:Prediction}, we address the problem of determining the hyperbolic response function parameters from information likely to be available to biologists in a typical microarray experiment, that is, fluorescence intensities for the complete set of features on each microarray, the probe sequences, and parameters associated with the experimental protocol.  We argue that information for the vertical scale parameters is in principal implicitly contained in the distribution of intensities across the microarray by partitioning the intensities by quantities which can be estimated from probe sequences such as probe-target binding energies, probe folding energies and probe pyrimidine content.  Determination of the horizontal scale parameter is a more formidable and open problem.  

Ultimately, our aim is find practical algorithms for biological analysis through an improved understanding of the physics of microarrays.  A problem encountered in in this paper, particularly with analysis of the horizontal scale parameter of the isotherm $K$, has been the difficulty encountered unravelling mutual correlations and compensations between competing effects.  In such cases one can never be totally certain, given the available data, that the physical interpretation is correct.  However, one could argue that is not necessary, nor possibly even helpful, to know all of the contributing physical effects in order to meet our ultimate aim.  If, for instance, our interpretation of the significance analysis of various models of the parameter $K$ in Section~\ref{sec:Horizontal} is correct, we discover that we often do just as well by modelling a number of physical effects by a single effective term.  In general, the lesson is that it is important to strike a balance between attempting to understand and incorporate all physical aspects on the one hand, and relying on empirically determined effective models on the other.  

The above analysis has concentrated on Affymetrix gene expression microarrays.  Genomic microarrays come in a variety of types for a variety of tasks.  As well as gene expression arrays there are tiling arrays, which interrogate large contiguous tracts of genome rather than specific genes, resequencing arrays designed to detect the location of mutations and single nucleotide polymorphisms (SNPs), SNP arrays which are used for genotyping, and non-DNA arrays such as protein arrays designed to identify protein-protein interactions and antibody arrays for detecting antigens.  In addition to this there exist a number of technologies including photolithographic deposition, inkjet printing and fabrication of probes onto micro-beads.  

In each case they share the common property that they detect large biological molecules of specific known letter sequences via their binding to complementary sequences attached to a solid surface.  In each case they will almost invariably share the common problems of non-specific hybridisation, saturation and bulk solution target hybridisation and folding dealt with in this work.  The physico-chemical model we have explored is consistent with spike-in data over a broad range of target concentrations and should serve as a starting point for a variety of microarray types and platforms.  

\section*{Acknowledgement} 

Thanks to Hans Binder, Susan Wilson and Yvonne Pittelkow for useful discussions and advice.  

\section*{Notation}
\begin{description}
\item[$a$:] {Physical background intensity measurement from factors such as reflection off the microarray surface and photomultiplier dark current.   Assumed to be constant for all features on a microarray.} 
\item[$A$:] {One of three parameters in the hyperbolic response curve Eq.~\ref{hyperbolicResponse} fitted to the measured fluorescence intensity data.  $A$ estimates the (background) fluorescence intensity at zero PM-specific spike-in concentration.} 
\item[$b$:] {Saturation fluorescence intensity above the physical background before washing, in a hypothetical situation in which all probes on a feature have formed biotin label carrying duplexes.  Assumed to be constant for all features on a microarray.}
\item[$B$:] {One of three parameters in the hyperbolic response curve Eq.~\ref{hyperbolicResponse} fitted to the measured fluorescence intensity data.  $A + B$ estimates the asymptotic saturation fluorescence intensity at infinite PM-specific spike-in concentration.} 
\item[$I(x):$] {Measured fluorescence intensity signal at PM-specific spike-in concentration $x$.}
\item[$K$:] {One of three parameters in the hyperbolic response curve Eq.~\ref{hyperbolicResponse} fitted to the measured fluorescence intensity data.  $K^{-1}$ estimates the PM-specific spike-in concentration required to give a fluorescence intensity half way between the background level $A$ and asymptotic level $A + B$.} 
\item[$s^\spec(t_{\rm W})$:] {The specific washing survival function, i.e. the probability that a duplex formed with a PM-feature-specific mRNA target  existing at the beginning of the washing step will survive to a washing time $t_{\rm W}$.}  
\item[$s^\NS(t_{\rm W})$:] {The non-specific washing survival function, i.e. the probability that a duplex formed with a PM-feature-non-specific mRNA target of species $i$ existing at the beginning of the washing step will survive to a washing time $t_{\rm W}$.}  
\item[$t_{\rm W}$:] The washing time.
\item[$x$:] {($= [S] + [S'] + \sum_i[S.NS_i]$) Spike-in concentration of mRNA PM-specific target.}
\item[$\Delta G^{\rm DNA/RNA}, \Delta G^{\rm RNA/RNA}, \Delta G^\Pfold$:] {Binding free energies of a DNA/RNA duplex, a RNA/RNA duplex and of DNA probe self-folding.  We use the convention that $\Delta G$ is negative for a bound state. $\Delta g^r$ are the corresponding dimensionless binding free energies $\Delta G^r/(RT)$, where $R$ is the gas constant and $T$ the absolute temperature.}   
\item[$\alpha$:]{The fraction of probes on the feature carrying probe-target duplexes after a washing time of $t_W$ at zero spike-in concentration $x$.  See Eq.~\ref{thetaIsotherm}.}
\item[$\beta$:]{$\alpha + \beta$ is the fraction of probes on the feature carrying probe-target duplexes after a washing time of $t_W$ at infinite spike-in concentration $x$.  See Eq.~\ref{thetaIsotherm}.}
\item[$\theta(x,t_W)$:] {The fraction of probes on the feature carrying probe-target duplexes after a washing time of $t_W$, as a result of a spike-in concentration $x$ of mRNA specific to the PM feature.  At $t_W = $ the end of the washing time, that is, at the time of scanning, we write simply $\theta(x)$.} 
\item[$\theta_\spec, \theta_\NS$:]{the fraction of probes on the feature carrying PM-specific and PM-nonspecific duplexes respectively at $t_W = 0$, i.e., after the hybridisation step and before the washing step.}  
\end{description}
\section*{Glossary}
\begin{description}
\item[{\it Hybridisation.}]{The reversible chemical reaction by which target molecules in solution bind to probes attached to the microarray surface to form duplexes.}  
\item[{\it Microarray.}]{A high-throughput device for detecting the presence of large biological molecules (DNA, RNA or proteins) of specific known letter sequences via their binding to molecules of complementary sequences attached to a solid surface.  They are high-throughput in the sense that large numbers of sequences are tested for in a single device.  The microarrays discussed here are oligonucleotide gene expression microarrays, that is, they have short DNA probes and are intended for the detection of expressed genes through their messenger RNA.}  
\item[{\it Non-specific hybridisation.}]{The hybridisation of target molecules with sequences other than those of the intended sequence.  Sometimes `non-specific' is used to mean `non-PM-specific', that is, hybridisation of target molecules   which are not complementary to the PM sequence, irrespective of whether they are binding to the PM or MM member of a probe pair.  PM and MM are defined in Section~\ref{sec:Datasets}.}
\item[{\it Probe.}]{A biological molecule attached to the microarray surface during fabrication.}
\item[{\it Spike-in experiment.}]{An experiment in which known concentrations of a specific set of target molecules are artificially added to a solution not otherwise containing those specific targets, and the solution hybridised onto microarrays.}
\item[{\it Target.}]{A biological molecule in the solution hybridised onto the microarray during a laboratory experiment.}  
\end{description}
\pagebreak

\end{document}